\documentclass[12pt,preprint]{aastex}

\slugcomment{to be submitted to The Astrophysical Journal}

\shorttitle{Carbon abundances in SMC PNe}
\shortauthors{Stanghellini et~al.}

\begin{document}

\title{Carbon abundance in Small Magellanic Cloud planetary nebulae through 
{\it Advanced Camera for Surveys} prism spectroscopy: constraining stellar evolution at low metallicity.}

\author{Letizia Stanghellini}
\affil{National Optical Astronomy Observatory, 950 N. Cherry Av.,
Tucson, AZ  85719}
\email{lstanghellini@noao.edu}

\author{Ting-Hui Lee}
\affil{Department of Physics and Astronomy, Western Kentucky University, 1906 College Heights Blvd $\#$11077, Bowling Green, KY 42101}
\email{ting-hui.lee@wku.edu}

\author{Richard A.~Shaw}
\affil{National Optical Astronomy Observatory, 950 N. Cherry Av.,
Tucson, AZ  85719}
\email{shaw@noao.edu}

\author{Bruce Balick}
\affil{Astronomy Department, Box 351580, University of Washington, Seattle WA 98195}
\email{balick@astro.washington.edu}

\and
\author{Eva Villaver}
\affil{Universidad Aut\'onoma de Madrid, Departamento de F\'isica Te\'orica C-XI, 28049 Madrid, Spain}
\email{eva.villaver@uam.es}

\begin{abstract}
We perform near ultraviolet ACS prism spectroscopy of 11 Small Magellanic Cloud (SMC) planetary nebulae (PNe) 
with the main aim of deriving the abundance of carbon. The analysis of the ACS spectra provide reliable atomic carbon abundances for all but a couple of our targets; ionic C$^{2+}$ abundances are calculated for all target PNe. 
With the present paper we more than double the number of SMC PNe with known carbon abundances, providing a good database to study the elemental evolution in low- and intermediate-mass stars at low metallicity. 
We study carbon abundances of Magellanic Cloud PNe in the framework of stellar evolution models and the elemental yields. Constraining SMC and LMC stellar evolutionary models is now possible with the present data, through the comparison of the final yields calculated and the CNO abundances observed. 
We found that SMC PNe are almost exclusively carbon rich, and that for the most part they have not undergone the hot-bottom burning phase, contrary to $\sim$half of the studied LMC PNe. The yields from stellar evolutionary models with LMC and SMC metallicities broadly agree with the observations. In particular, evolutionary yields for M$_{\rm to}<3.5~{\rm M}_{\odot}$ well encompass the abundances of round and elliptical PNe in the SMC. We found that the carbon emission lines are major coolants for SMC PNe, more so than in their LMC counterparts, indicating that metallicity has an effect on the physics of PNe, as predicted by \citet{stanghellini03}.

\end{abstract}


\section{Introduction}

Planetary Nebulae (PNe) have been studied for decades, and through those 
efforts an understanding has come of 
the final phases of stellar evolution of stars with masses in 
the $\sim$1 -- 8 M$_{\odot}$ range (also called the low- and intermediate-mass stars, or LIMS).  
Planetary nebulae are ejected toward the end of the LIMS evolution, during the final thermal pulses
on the Asymptotic Giant Branch (TP-AGB), thus they are 
ideal probes to test the ISM enrichment from these stars in a quantitative way. 
The scientific importance of studying 
PNe in the Magellanic Clouds (LMC, SMC) can be
very simply summarized: LMC and SMC PNe are absolute probes of 
stellar and nebular brightness, size, and shape, because their distance
is relatively well known,
and because their interstellar extinction is comparatively low and 
uniform. In particular, 
the SMC is a unique laboratory to study
resolved stellar populations in a low metallicity environment:
chemical analysis of the PNe therein gives a direct test of cosmic recycling at low metallicity, 
which is essential to understanding the recycling of elements within galaxies at high redshift.

The gas ejected at the PN phase
contains elements that have been produced in the stellar interiors, and then 
carried to the stellar surface by the convective dredge-up processes 
(Iben \& Renzini 1983, van den Hoek \& Groenewegen 1997).
The single-star evolution models predict a chemical enrichment of the 
outer region of evolved LIMS, both for Galactic and
Magellanic Cloud LIMS alike, which can be 
summarized as follows: 

During the first Red Giant phase, the convective envelope penetrates
regions that are partially CNO-processed. This dredge-up results in a $^{13}$C 
and $^{14}$N enhancement, and a decrease of $^{12}$C.
Afterward, He-burning starts in the stellar core.  
Later, the hydrogen and helium burning occur alternately 
in two nuclear burning shells, surrounding the CO core, and the star
ramps up to the AGB.

The second convective dredge-up (for M$_{\rm to}$, the {\it turnoff} mass, larger 
than $\sim$3 M$_{\odot}$) occurs at the onset of the AGB phase,
when the H-burning shell is temporarily extinguished. This process carries $^4$He, $^{13}$C,
and $^{14}$N-rich material to the stellar surface. 

During the TP-AGB phase, the envelope is able to dredge-up material 
after each thermal pulse, carrying $^4$He, $^{12}$C, and other relatively
light s-process elements to the surface. This process is called the third dredge-up.

For M$_{\rm to}>$3--5 M$_{\odot}$ (exact mass depends on metallicity, Marigo
2001), an additional process 
is thought to occur that alters the chemical composition: the so-called 
hot-bottom burning (HBB) that processes most of the carbon into nitrogen,
occurring during the quiescent interpulse periods between thermal pulses.

The key to assessing the above predictions is to measure the abundances 
of the processed elements, particularly C, N, and O. 
Carbon and nitrogen enrichment depends on the progenitor mass, yielding to
a direct connection between
observed progenitor mass (i.~e., population) and chemical content. By measuring 
the C and N abundances in PNe, one can 
at once validate key elements of stellar evolution theory, and 
measure the contribution of LIMS to the enrichment of the ISM.  While O and N abundance
analysis is straightforward 
to do with PNe, owing to their bright optical emission lines, the carbon analysis requires satellite UV spectroscopy.

In our earlier study of LMC PN abundances (Stanghellini et al. 2005)
we have shown that PN morphology  is a surprisingly useful indicator of 
the progenitor stellar evolution and population.
We used a subsample of LMC PNe images available from {\it HST} 
\footnote{LMC and SMC PNe are typically
less than 0.5 arcsec across, they can be physically resolved 
from space. } and measured the carbon abundances with STIS spectroscopic
analysis. We found that nitrogen 
enhancement is correlated with asymmetry.
These results are consistent with the predictions of stellar evolution 
only if the progenitors of asymmetric PNe have on average larger masses than 
the progenitors of symmetric PNe. 
Our results are the first of the kind for extra-galactic PNe, and are
thus not biased by the
large selection effects that limit the observation of PNe in the Galactic
disk.

Owing to the smaller metal abundance in the SMC than in the LMC, it is very important to extend  the LMC study
to the SMC. Carbon abundance determination in SMC PNe would probe the metal enrichment rates for low
metallicity stelar population.

On these grounds we obtained near UV spectra of 11 SMC PNe that, together with the already published 
LMC and SMC UV data, provide a database that has a statistical impact on the
study of carbon in Magellanic Cloud PNe. In the present paper we discuss the 
data acquisition and analysis of the ACS prism spectra ($\S$2), the abundance analysis to derive
carbon in SMC PNe ($\S$3), and a discussion of our results in the framework on PN evolution and populations ($\S$4). 
The conclusions are given in $\S$5.

\section{Observations and Data Analysis}
\subsection{Observations}

We observed 11 SMC PNe using the Advanced Camera for Surveys (ACS)
prisms PR130L in the Solar Blind Channel (SBC) and PR200L in the High
Resolution Channel (HRC). Our aim is to detect the \ion{C}{4} $\lambda$1550, 
the \ion{C}{3}] $\lambda$ 1909, and the \ion{C}{2}] $\lambda$2326 nebular emission
lines, to determine the ionic abundances of the carbon ions that 
are excited in the PN regime. 

We select our targets from the SMC PNe that have been
previously observed by us with the {\it Space Telescope Imaging Spectrograph}
(STIS, Stanghellini et al. 2003; Shaw et al. 2006). With this selection we guarantee that
their  
morphology, size, optical extinction, and optical fluxes are known.
From the \citet{stanghellini03, shawetal06} sample we chose all PNe that we deemed bright enough 
to be observed with the observing configuration within a few orbits (total
H$\beta$ fluxes larger than 2.5$\times 10^{-14}$ [erg cm$^{-2}$ s$^{-1}$]).  
In addition, our selected targets have
angular sizes smaller than  $\lesssim 0.5''$ in order
to prevent blending of emission lines in the slitless spectroscopy.
The targets are listed in Table \ref{tb:log}, together with the observing log.
Of these, a few have been observed before with the IUE, but none has a sound carbon abundance
determination.

The SBC detector is a $1024 \times 1024$
solar-blind CsI Multi-Anode Microchannel Array (MAMA), with $\sim
0.034 \times 0.030$\arcsec~~~~ pixels, and a nominal $35 \times 31$\arcsec field of view.  
The HRC detector is a $1024 \times 1024$
SITe CCD with $\sim 0.028 \times 0.025$\arcsec pixels, covering a
nominal $29 \times 26$\arcsec field of view.  The wavelength range
of SBC/PR130L is $\sim$ 1200- 2000 \AA, and the useful wavelength
range of HRC/PR200L is $\sim$ 1800- 4000 \AA.  The wavelength
scale of the prisms is non-linear, with spectral resolution decreasing
towards longer wavelengths.  For PR130L, the dispersion varies from
about 2 \AA/pixel at the blue end (R $\sim$ 300), to $\sim$ 10
\AA/pixel at 1600 \AA~(R $\sim$ 80) and 30 \AA/pixel at the
red end (R $\sim$ 30 at 2000\AA).  For PR200L, the dispersion varies
from about 6 \AA/pixel at the blue end (R $\sim$ 150), to $\sim$
20 \AA/pixel at 2500 \AA~(R $\sim$ 60) and $\sim$ 200 \AA/pixel at 
4000 \AA~(R $\sim$ 10) \citep{larsenwalsh06, larsen06}.

For each target, the HRC observations consist of a direct image through a
broad band filter
(F330W), and two exposures through the prism (PR200L) in order to perform rejection of
cosmic rays. Similarly, the SBC observations consist of a direct
image through F165LP, and one exposure of PR130L, 
unless the exposure time
is longer than one orbit, in which case two exposures were taken.  The
direct images have been acquired to establish the zero-point of the wavelength
scale.

\subsection{Data Analysis}

The pipeline calibration of our data provided bias-subtracted, dark-corrected, 
and flat-corrected spectral images, as processed with CALACS \citep{pavlovsky04}.
We used MultiDrizzle and aXe to combine the images, correct for geometric distortion,
and to identify any bad pixels in the spectrum. The bad pixels were excluded from the 
analysis of our flux-calibrated one-dimensional spectra.

The position, size, and magnitude of the extracting sources were
identified and determined using the direct images SBC/F165LP and
HRC/F330W via analysis with SExtractor \citep{bertin96}. 
The position of the extracting sources on the 
direct images have then been  projected on the prism images
for spectral extraction.  

The spectral extraction was done with the aXe software \citep{kummel05}
in PyRAF\footnote{PyRAF is a product of the Space Telescope Science
Institute, which is operated by AURA for NASA.}.  For prism slitless
spectroscopy, the trace and wavelength solutions include spatial
variations across the HRC and SBC detectors.  The wavelength and flux
calibrations are provided by the ST-ECF group in the configuration
files for the aXe software.  Two white dwarf standards were used to
determine the flux calibration for both HRC and SBC prisms.  The
HRC/PR200L wavelength calibration has been secured with observations of
the LMC PN SMP-79
and a quasar with both STIS and the ACS prism \citep{larsenwalsh06}. 
The calibration of the  SBC/PR130L prism was
determined using two quasars \citep{larsen06}.

Background subtraction was performed during spectral extraction.  To
remove the sky background, a local background is estimated by
interpolating between the adjacent pixels on either side of the target
spectrum, outside the extracting area. The optimal weighting algorithm was
chosen to enhance the signal-to-noise
ratio of the extracted spectra. The algorithm assigns lower weights to pixels which contain
only a small fraction of the target flux, due to the larger distance
from the spectral trace.

Faint objects near the targets were
identified by position and size, and then their contribution to the flux was subtracted
in the extracted spectra. The contaminating flux from
all other sources was estimated with a gaussian emission model by
using the sizes and magnitudes derived from the direct image.  This
contamination flux was then subtracted when extracting the target
spectrum.

\begin{figure}
\plottwo{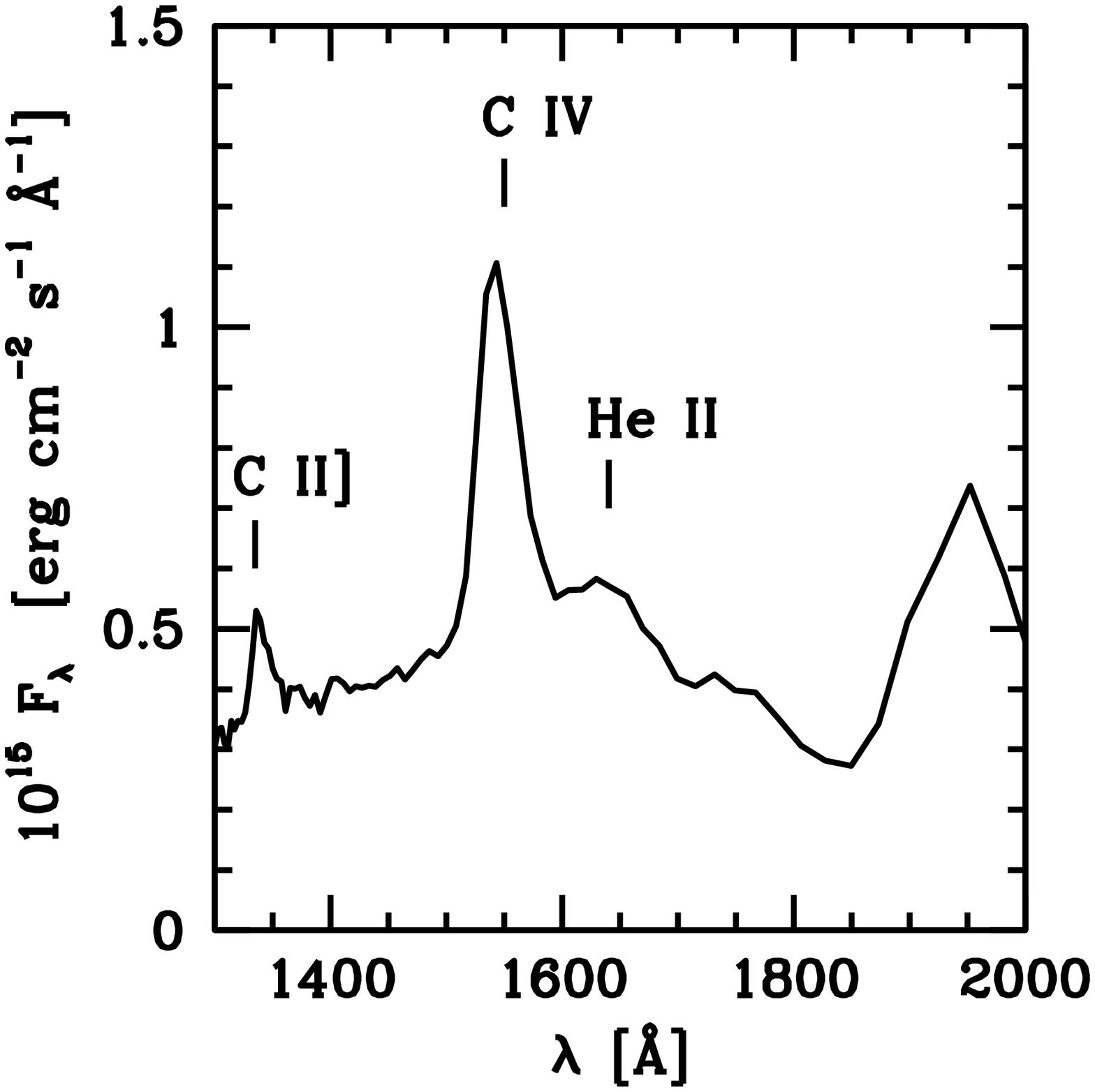}{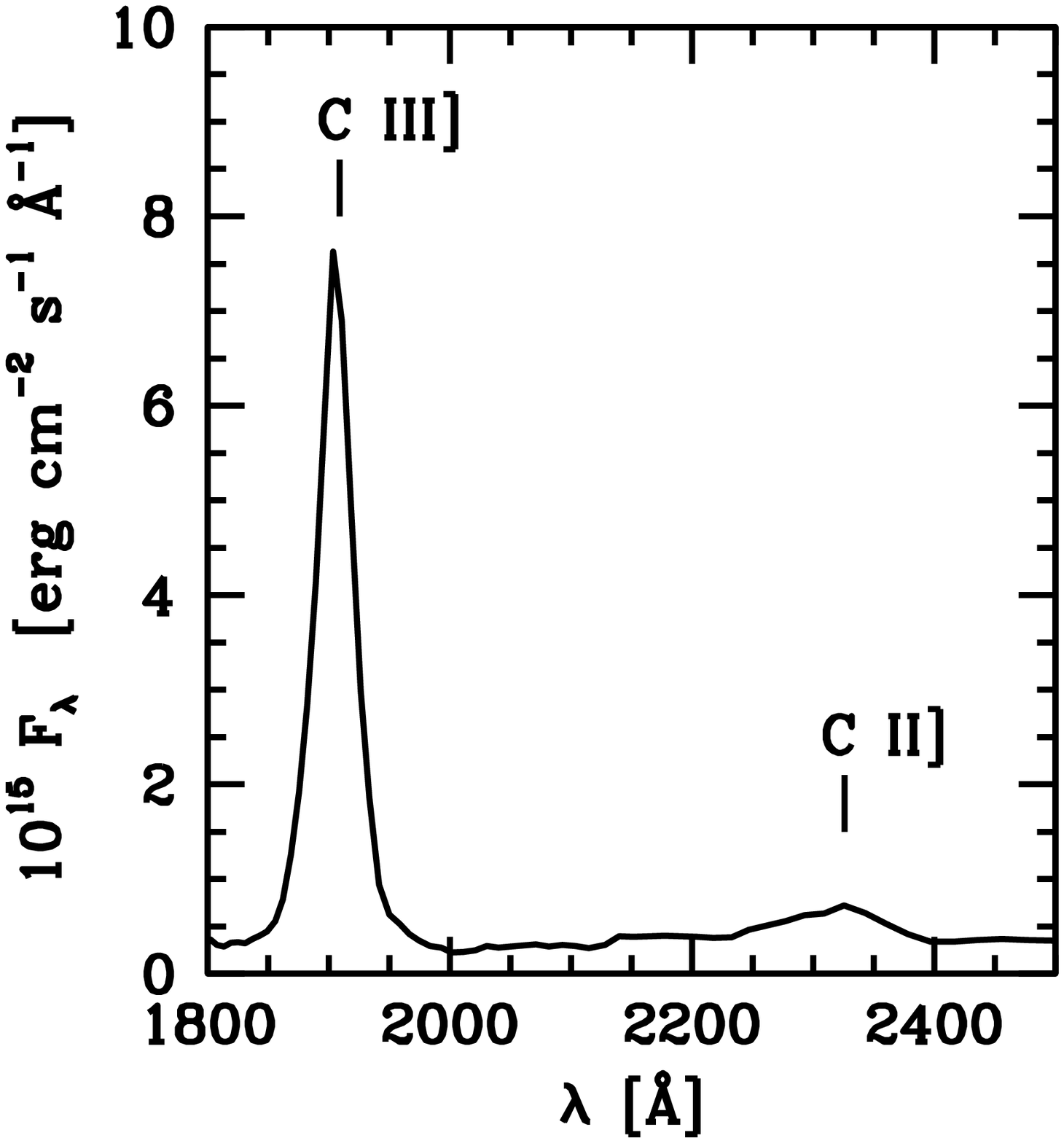}
\caption{The extracted spectra from the PR130L (left panel) and PR200L (right panel)
prism observations in SMP~6.}
\label{fg:smp6}
\end{figure}

\begin{figure}
\plottwo{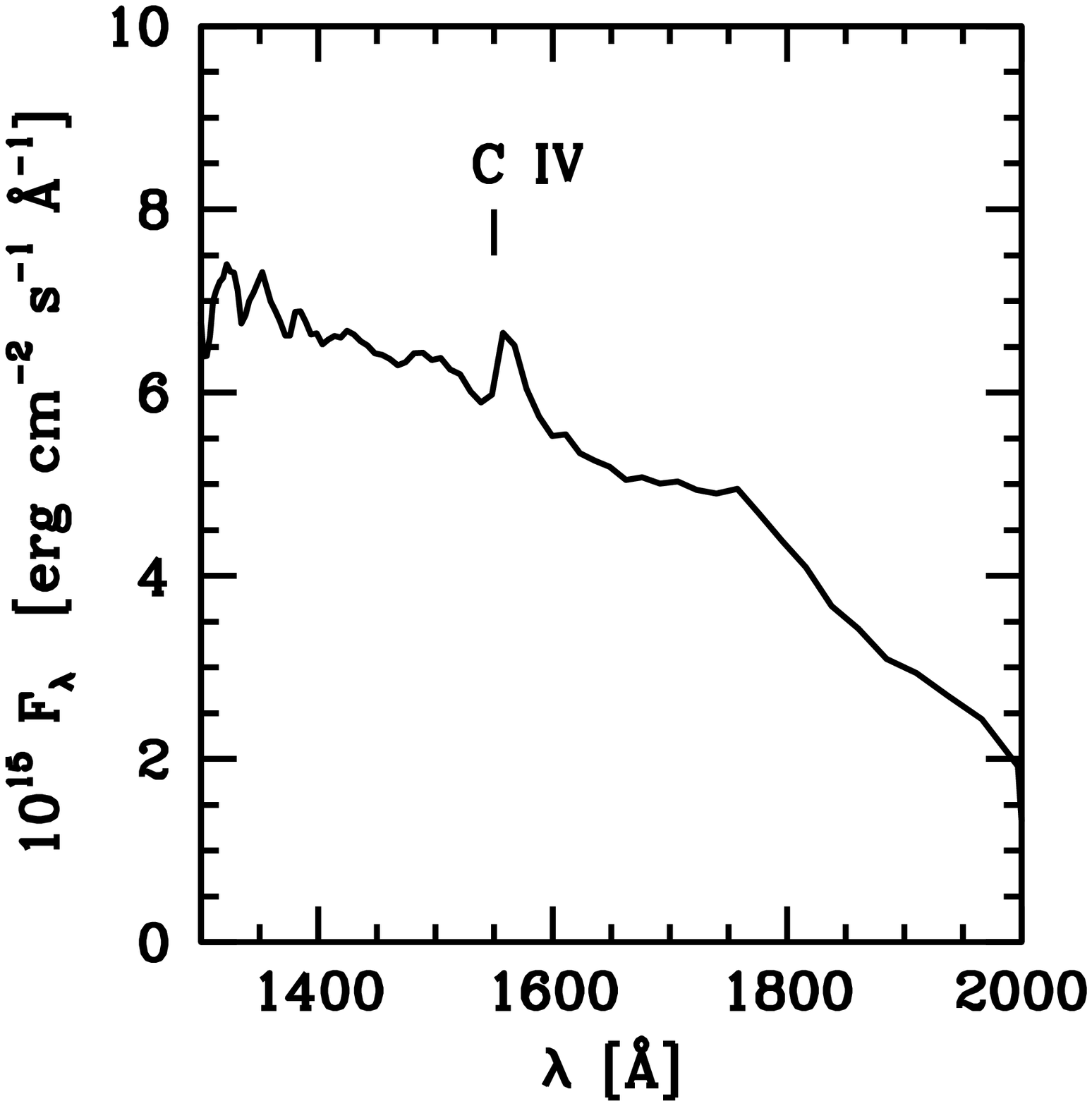}{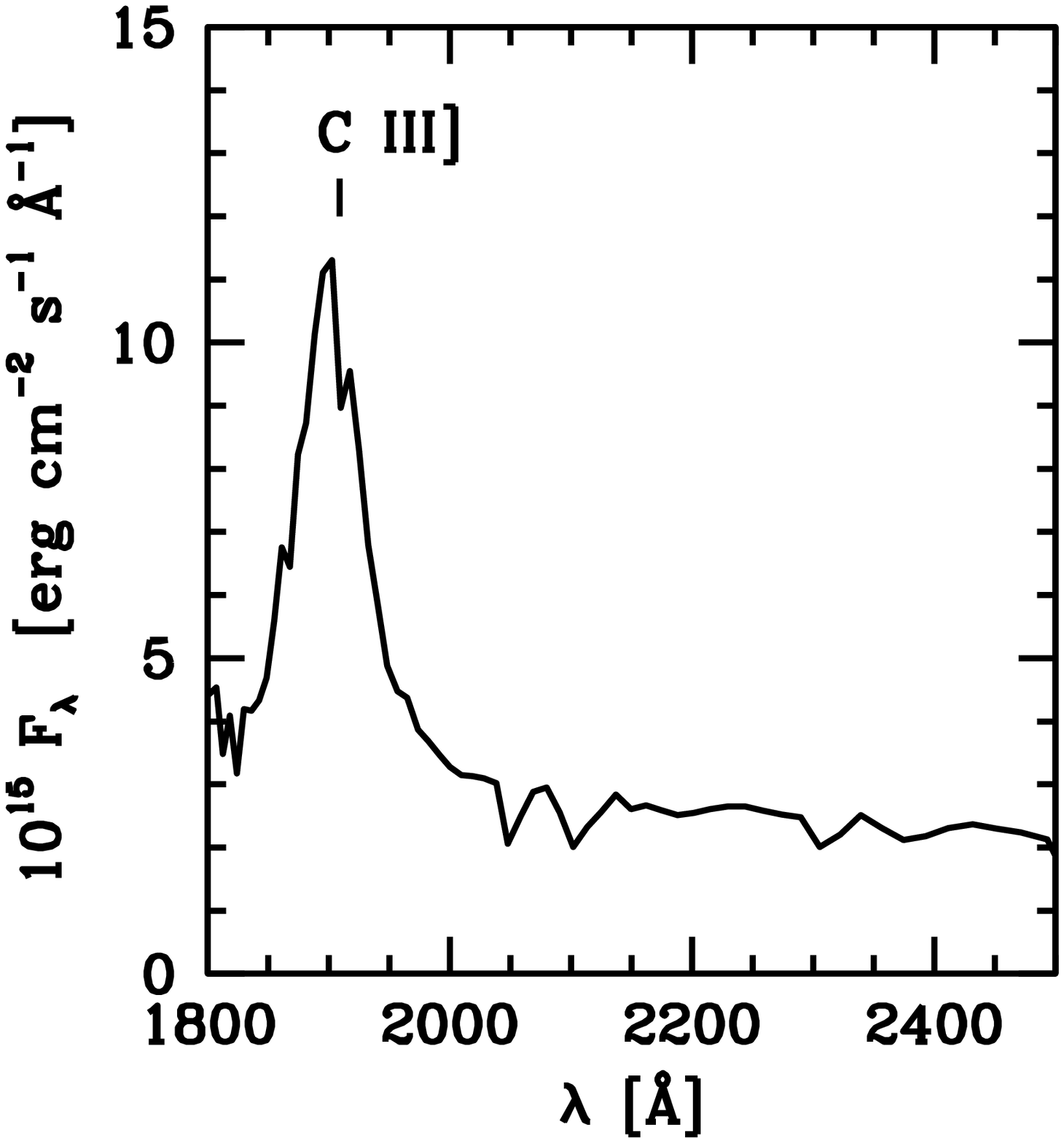}
\caption{As in Figure 1, for SMP~8.}
\label{fg:smp8}
\end{figure}

\begin{figure}
\plottwo{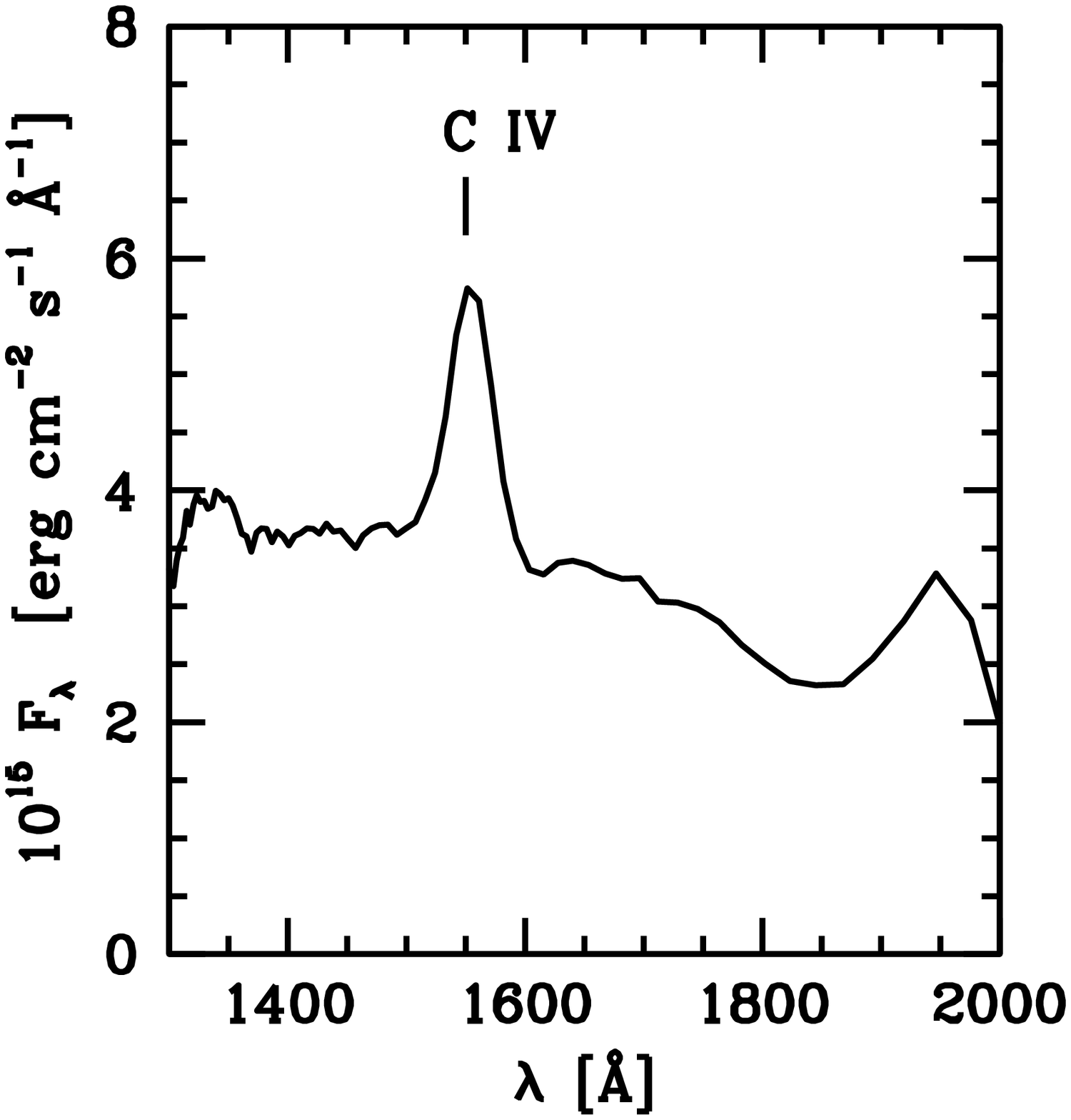}{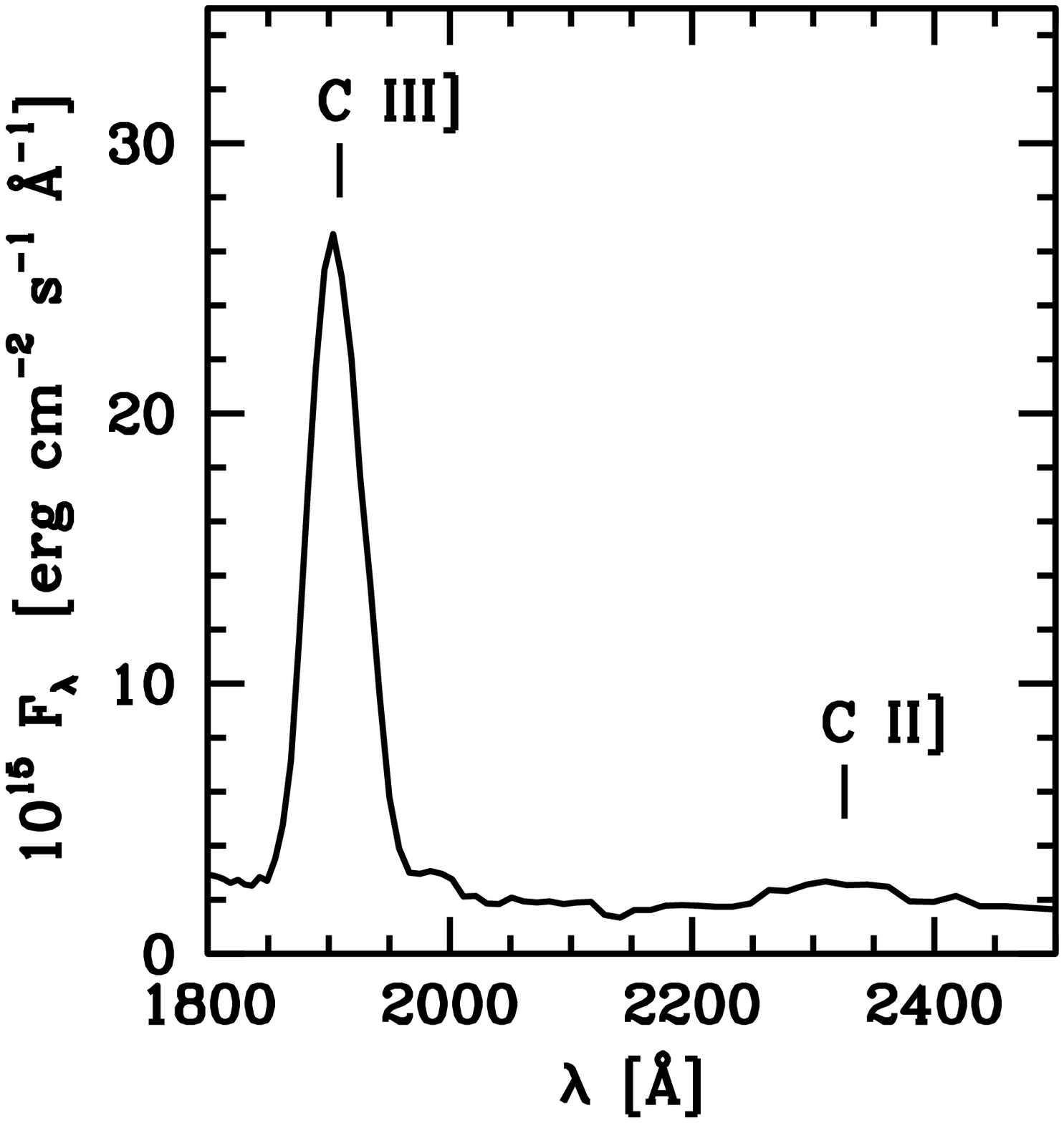}
\caption{As in Figure 1, for SMP~13.}
\label{fg:smp13}
\end{figure}

\begin{figure}
\plottwo{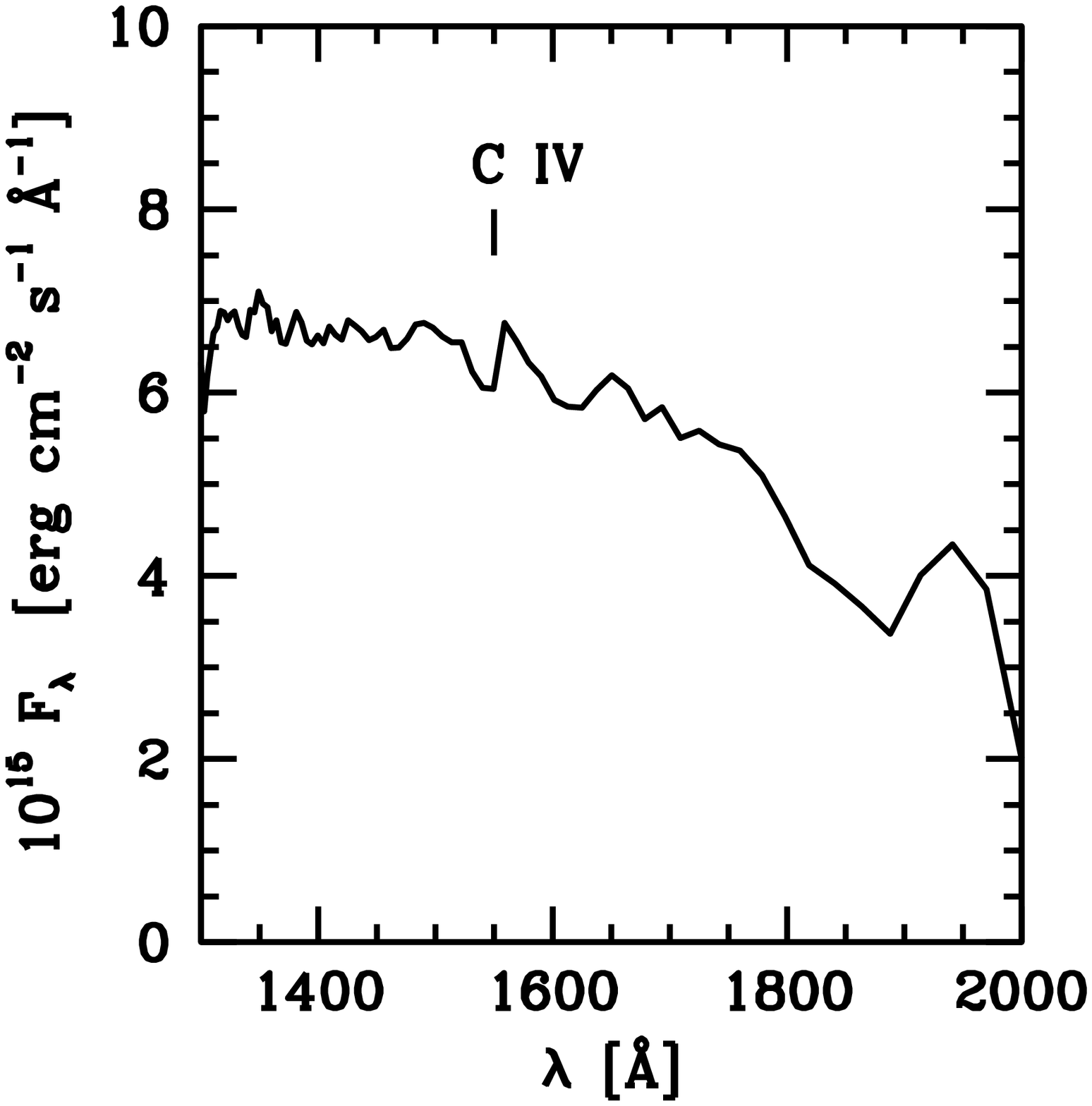}{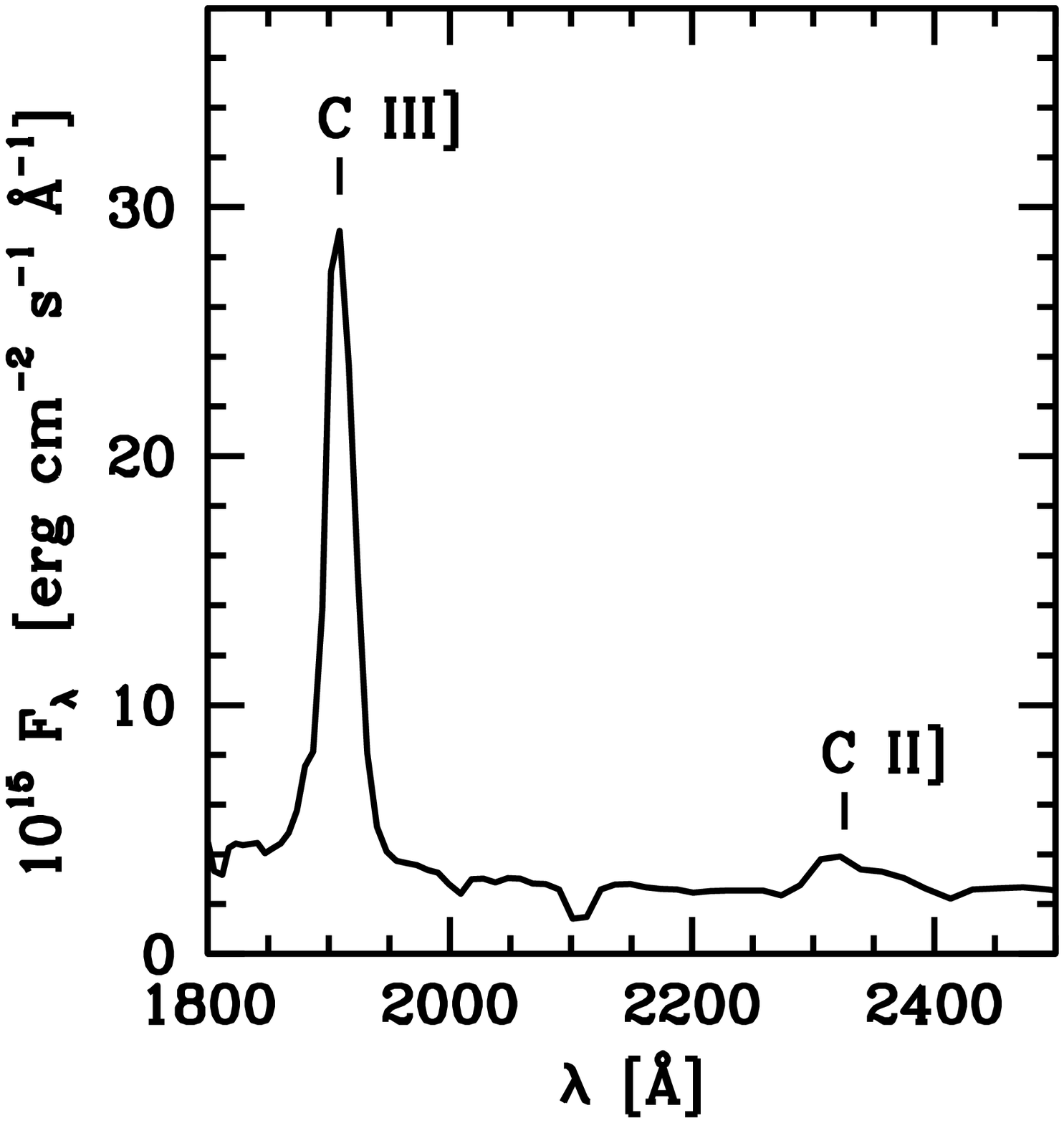}
\caption{As in Figure 1, for SMP~15.}
\label{fg:smp15}
\end{figure}

\begin{figure}
\plottwo{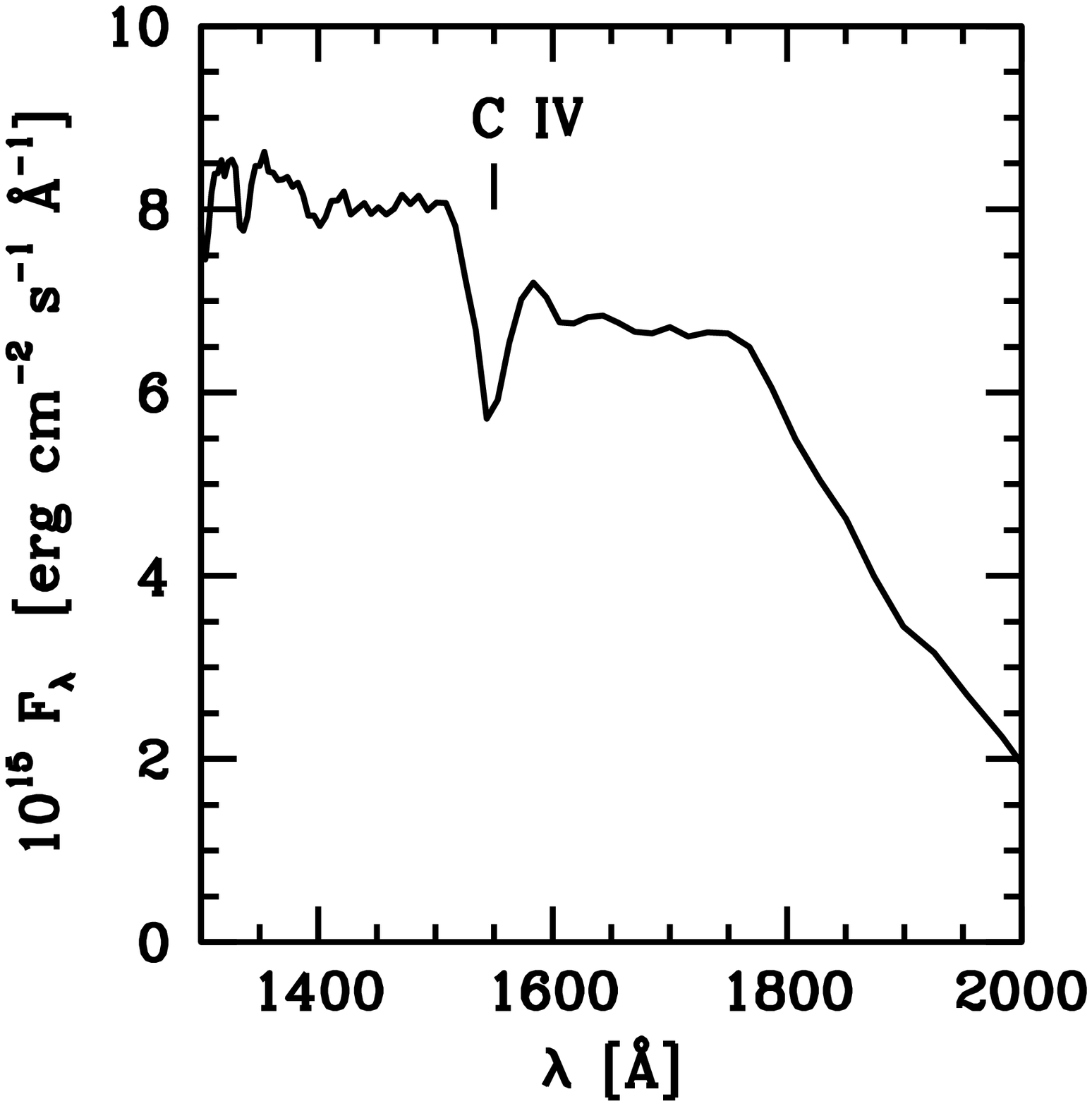}{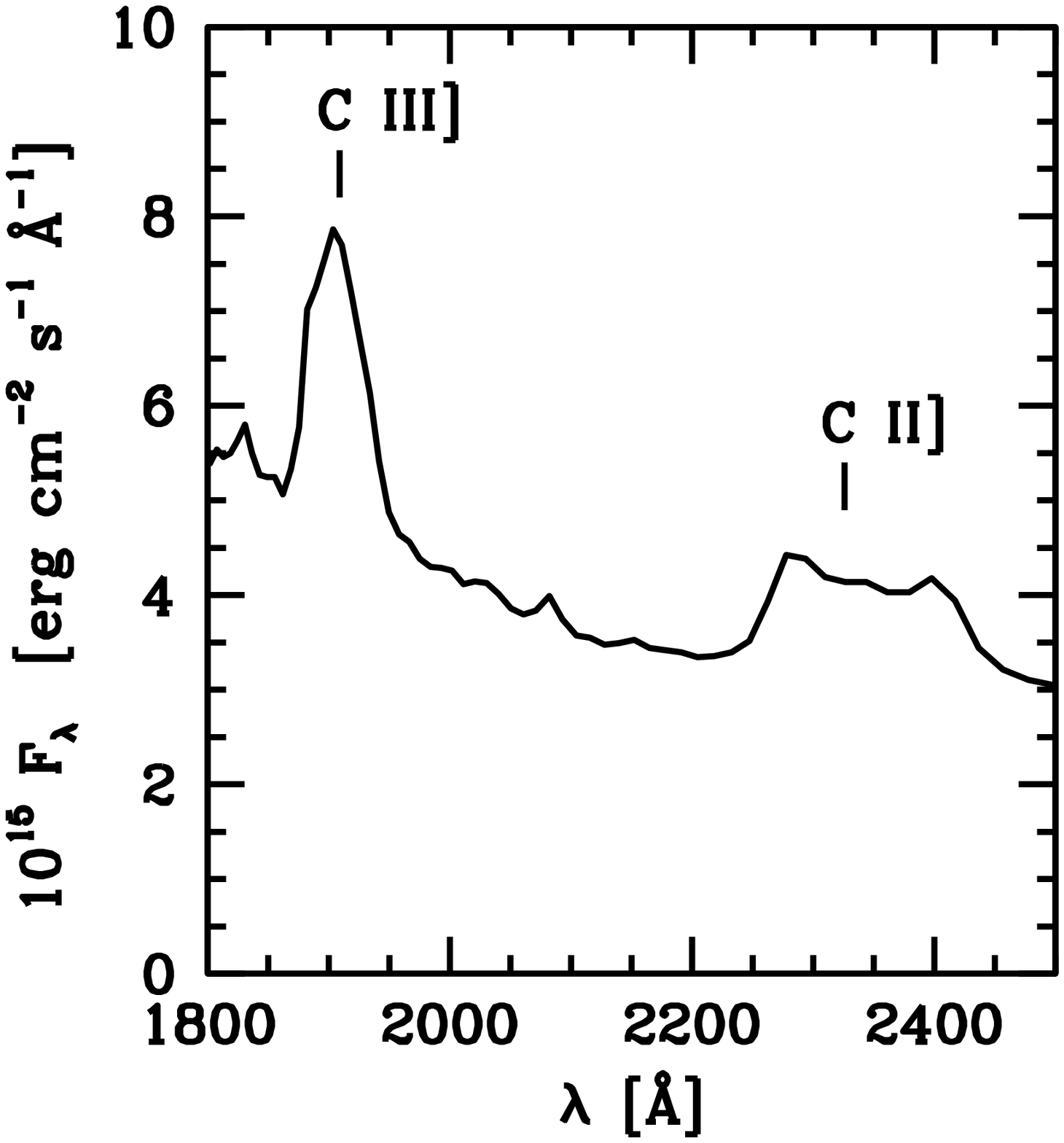}
\caption{As in Figure 1, for SMP~16.}
\label{fg:smp16}
\end{figure}

\begin{figure}
\plottwo{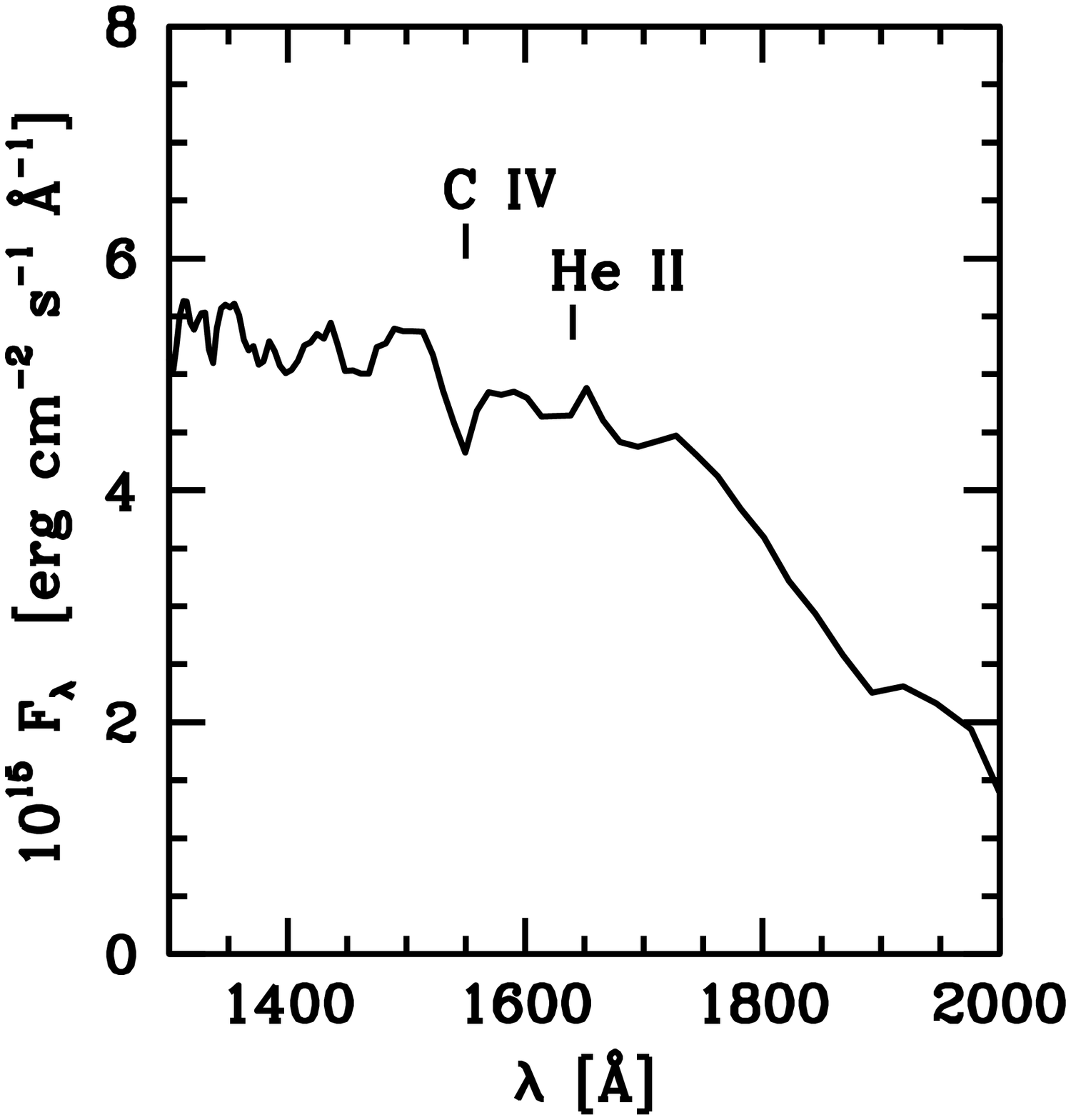}{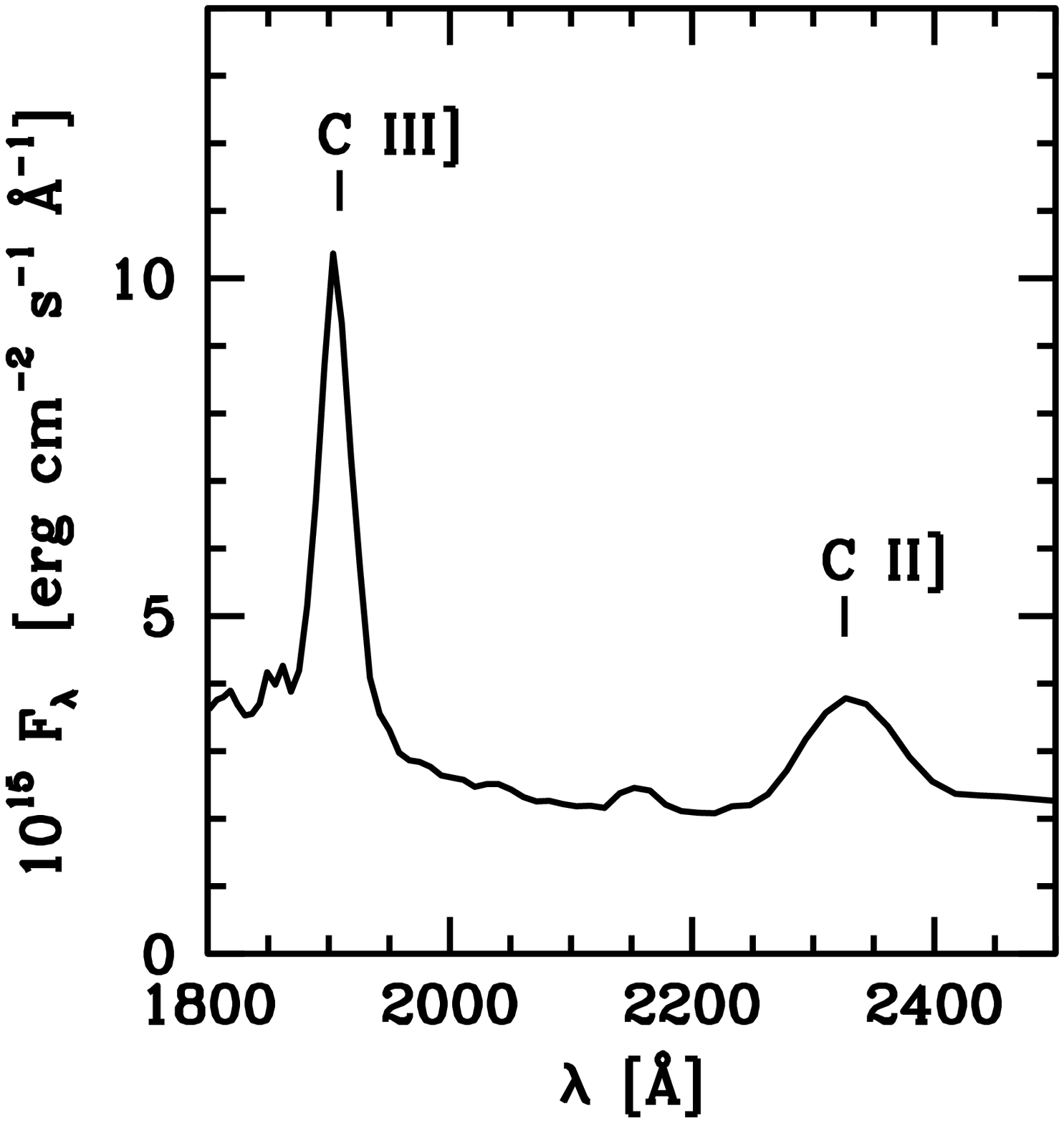}
\caption{As in Figure 1, for SMP~18.}
\label{fg:smp18}
\end{figure}

\begin{figure}
\plottwo{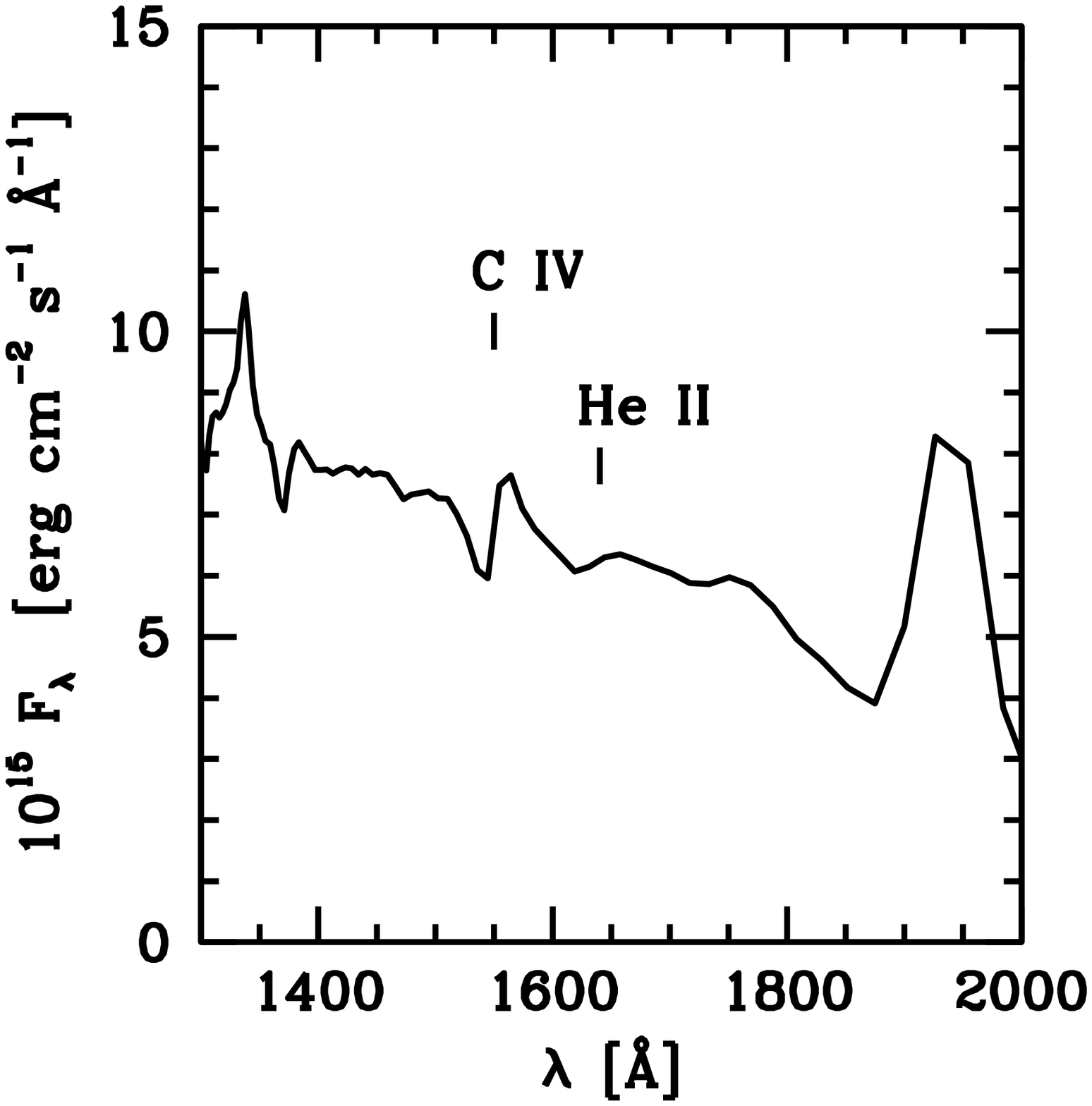}{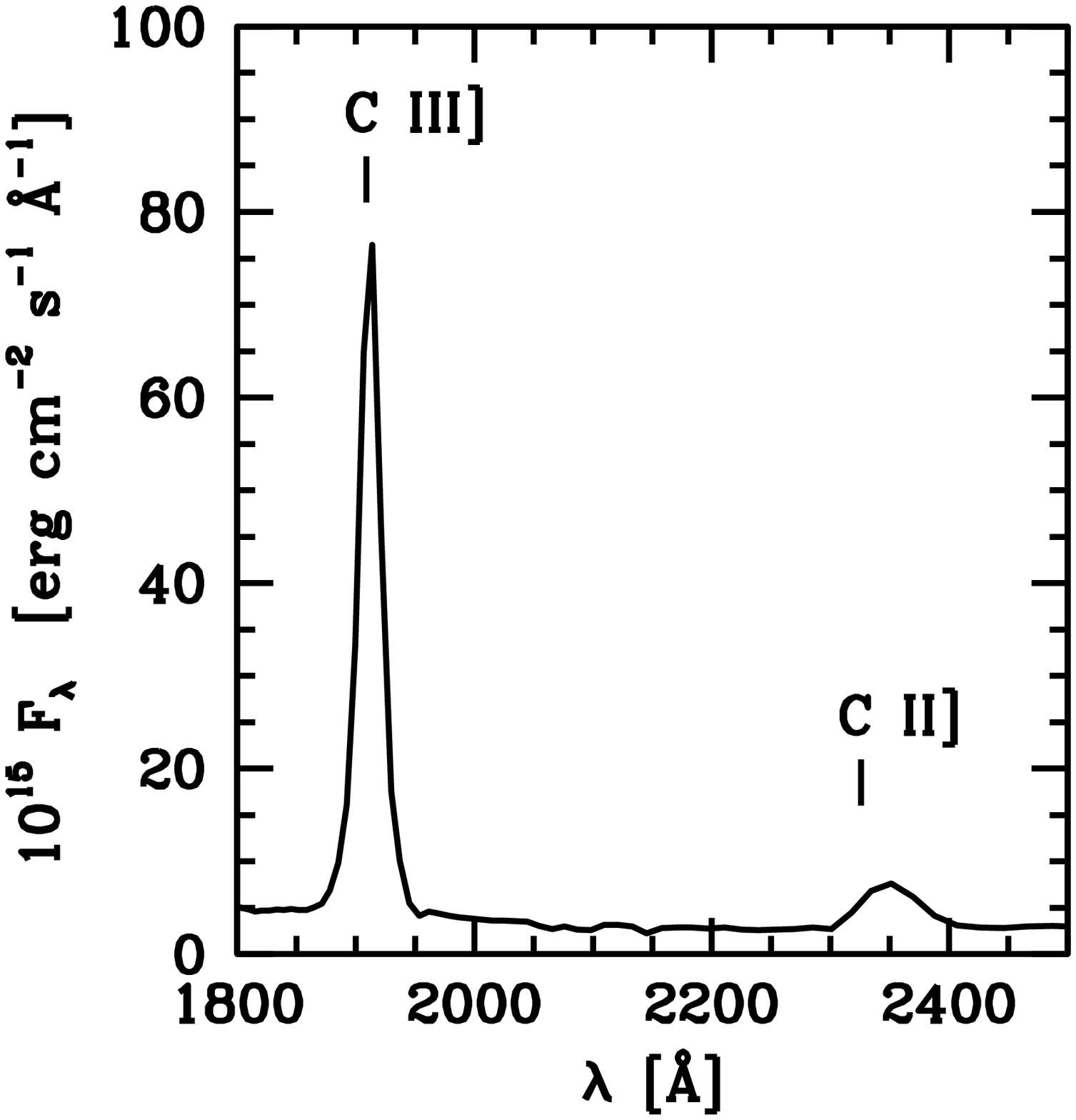}
\caption{As in Figure 1, for SMP~20.}
\label{fg:smp20}
\end{figure}

\begin{figure}
\plottwo{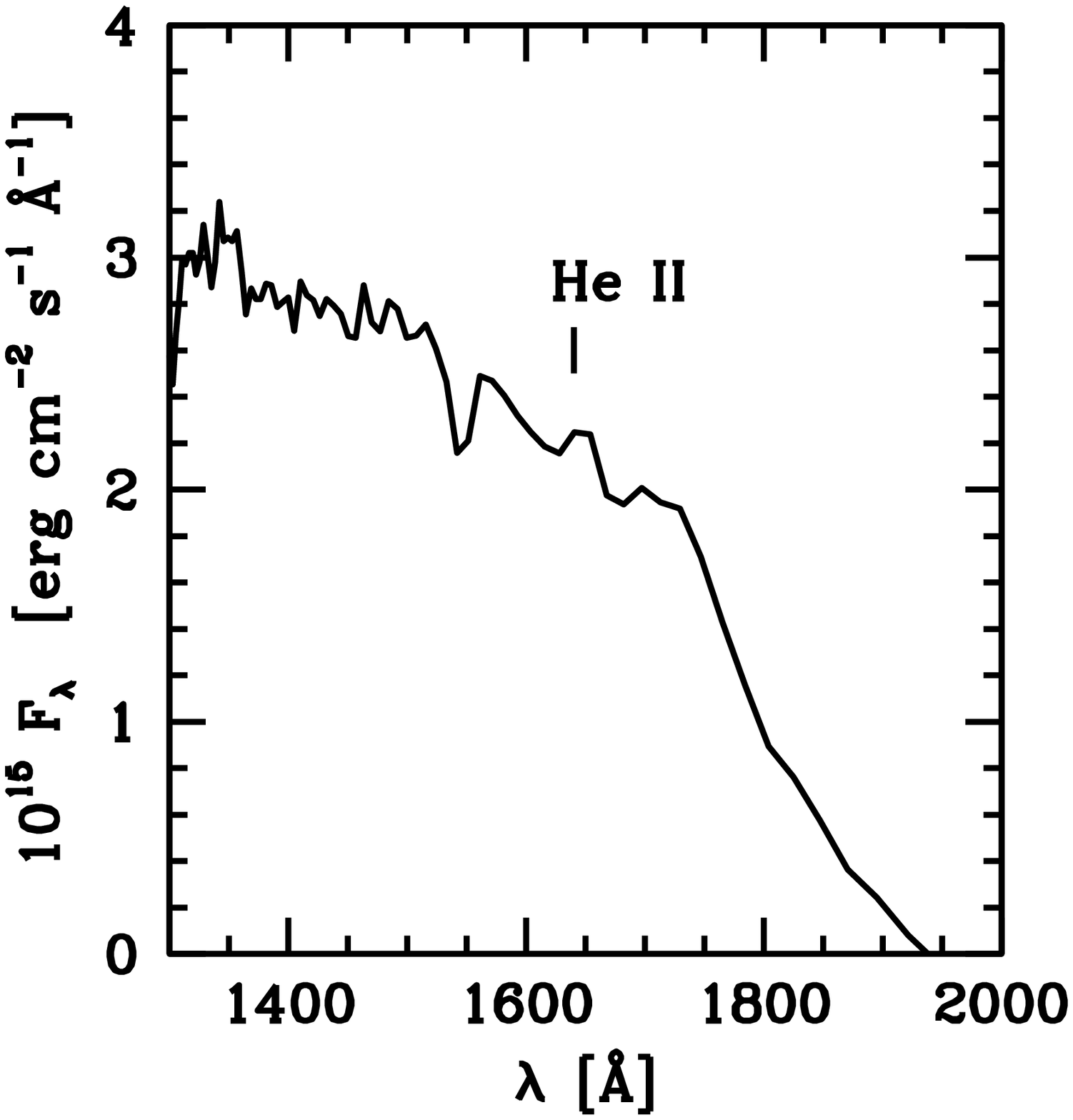}{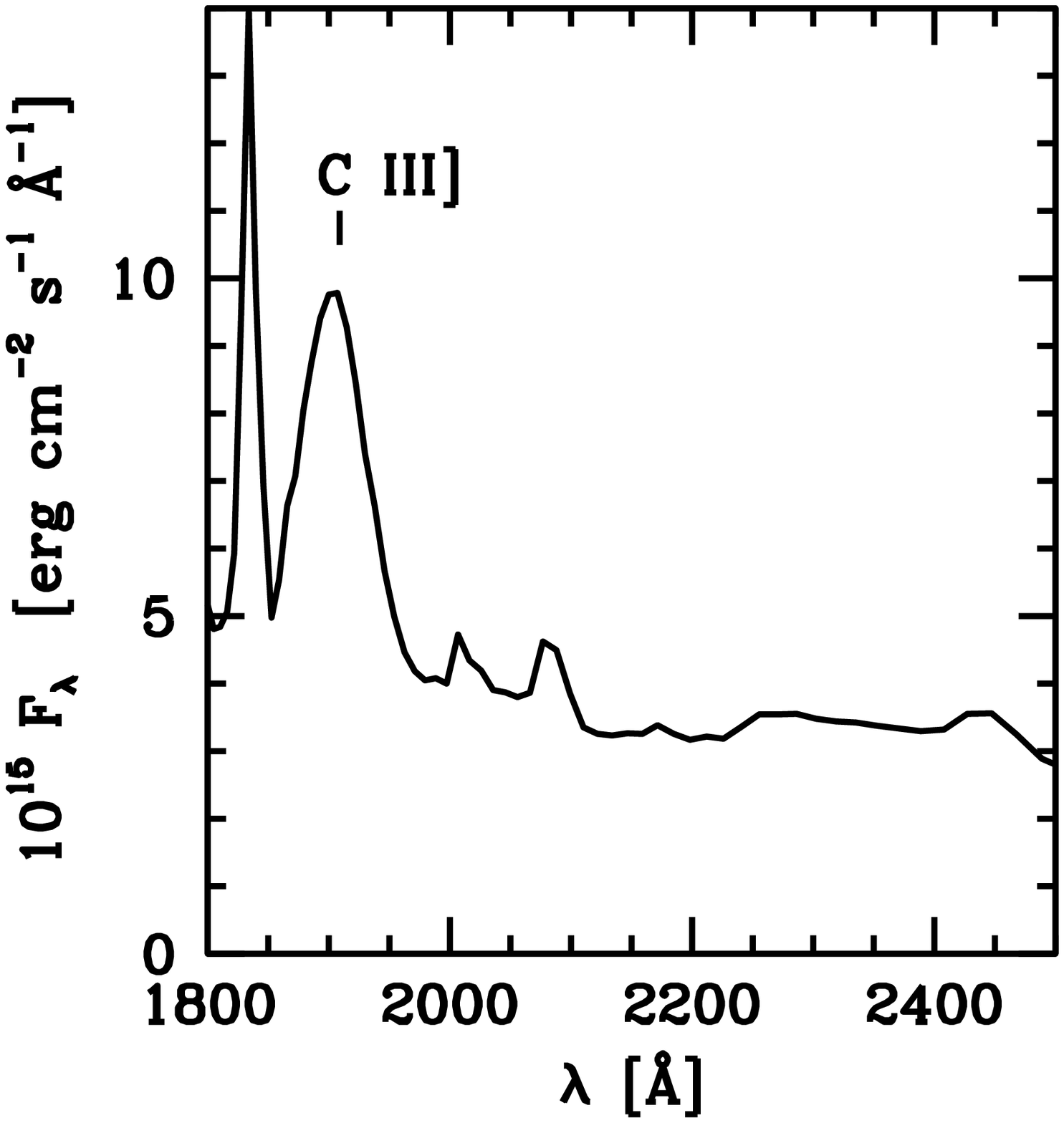}
\caption{As in Figure 1, for SMP~24.}
\label{fg:smp24}
\end{figure}

\begin{figure}
\plottwo{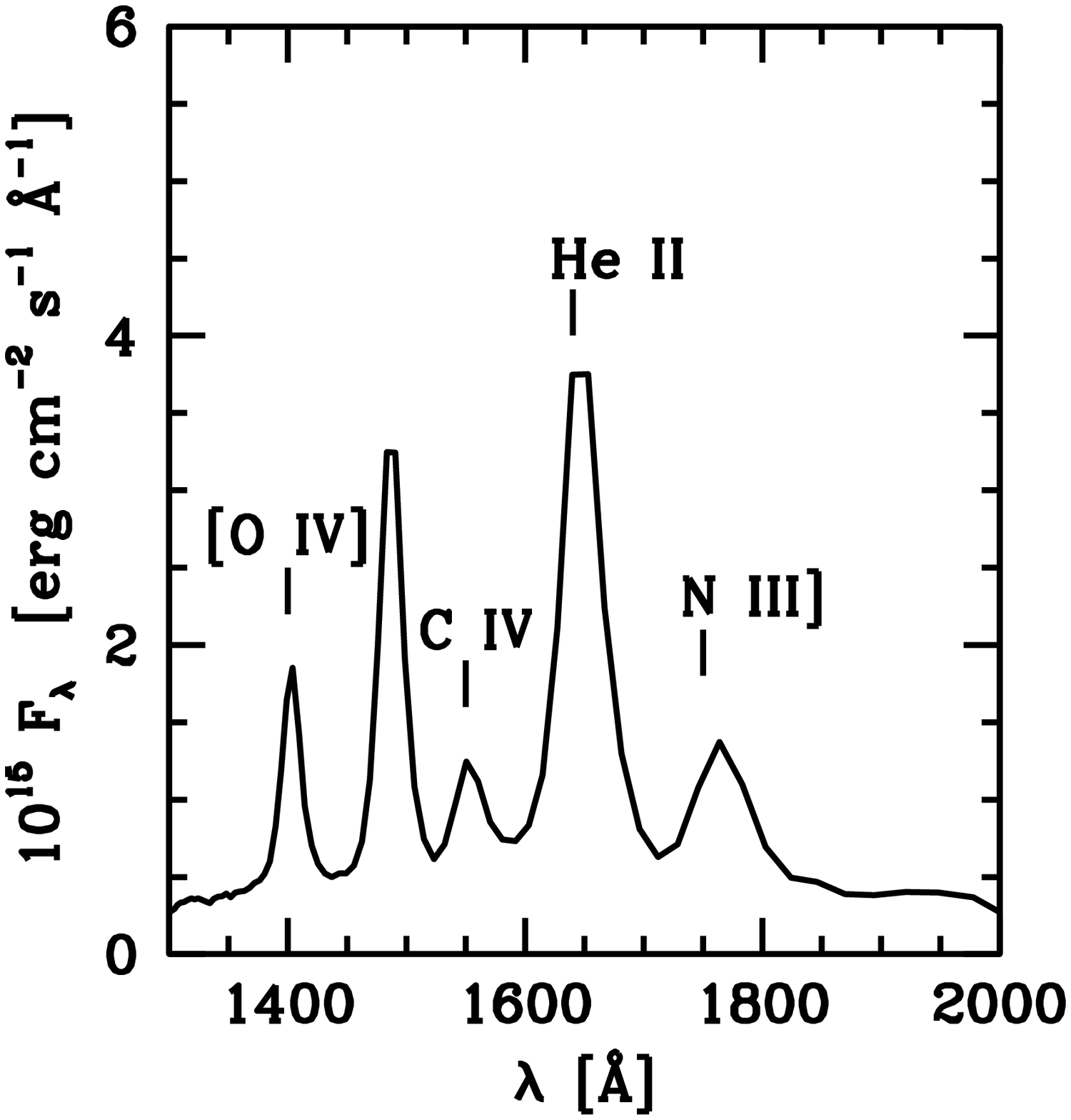}{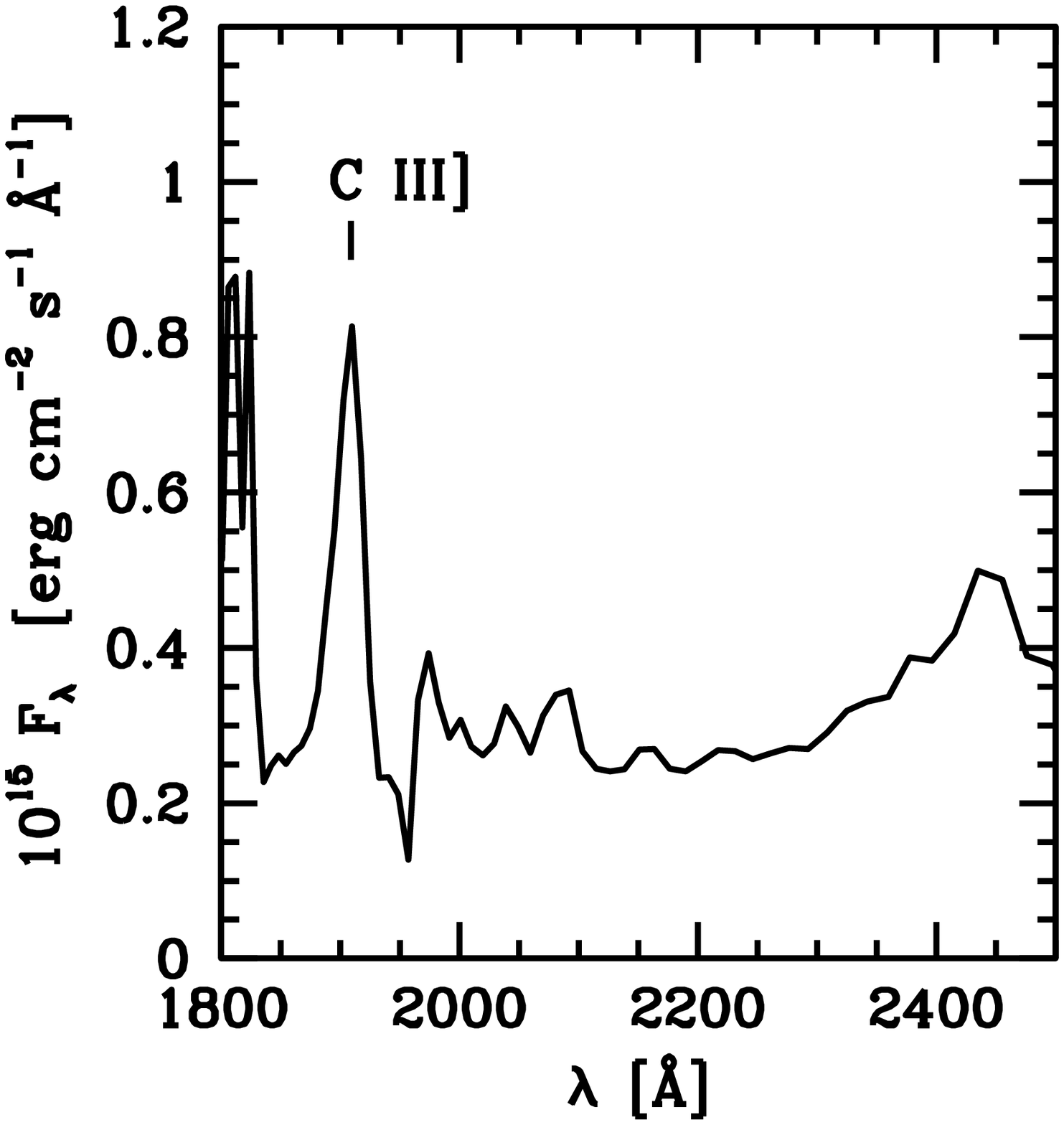}
\caption{As in Figure 1, for SMP~25.}
\label{fg:smp25}
\end{figure}

\begin{figure}
\plottwo{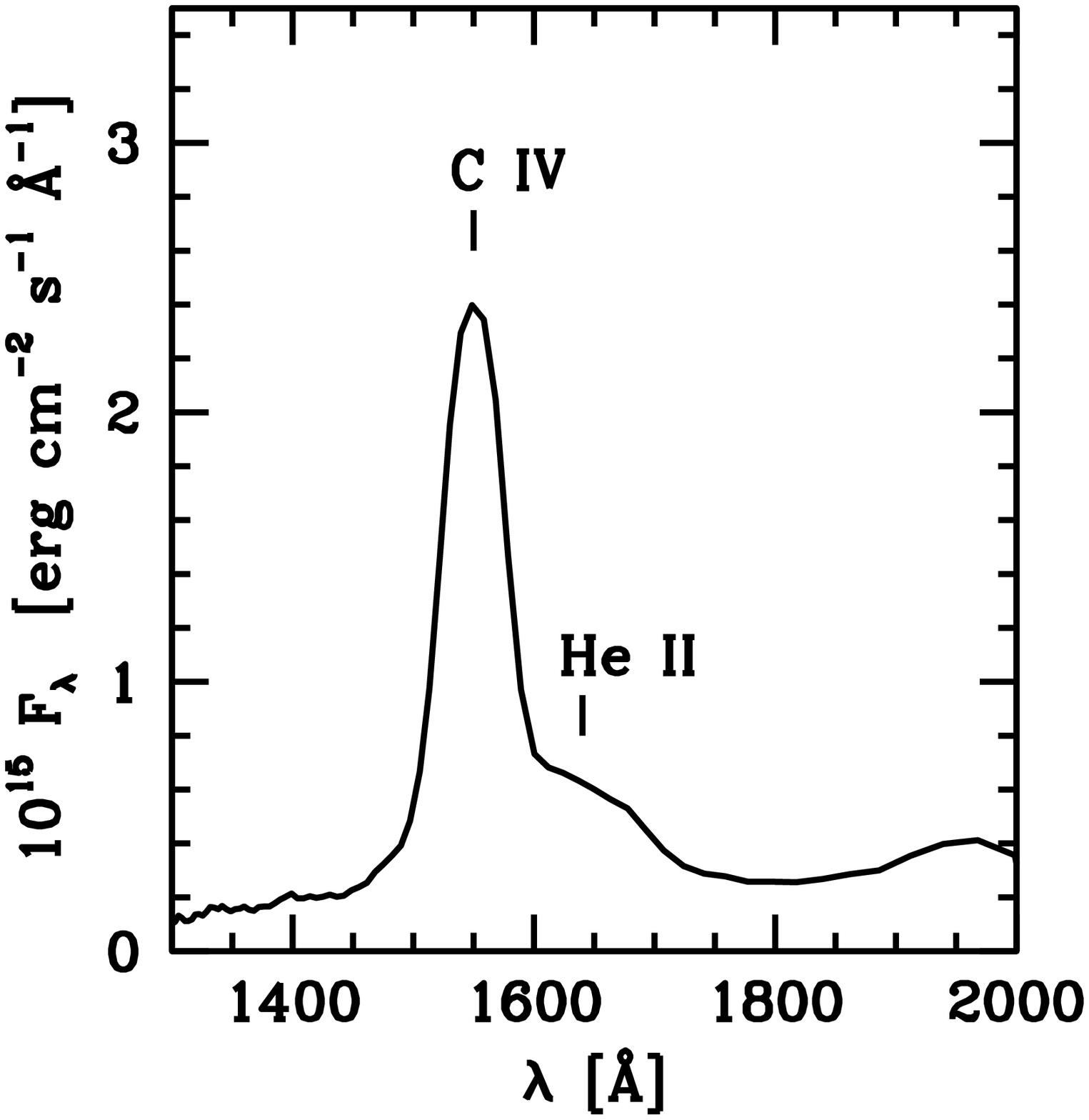}{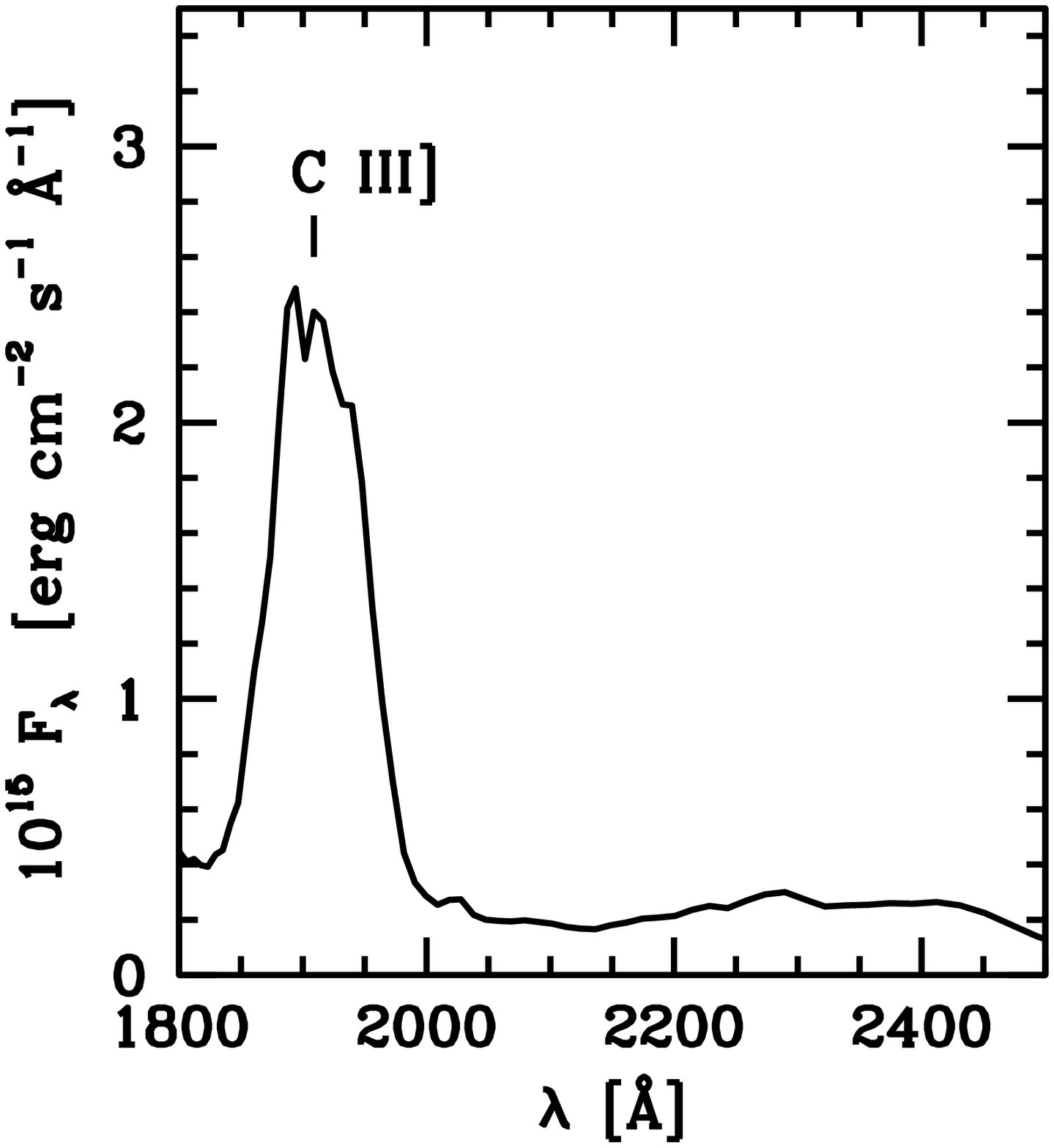}
\caption{As in Figure 1, for SMP~26.}
\label{fg:smp26}
\end{figure}

\begin{figure}
\plottwo{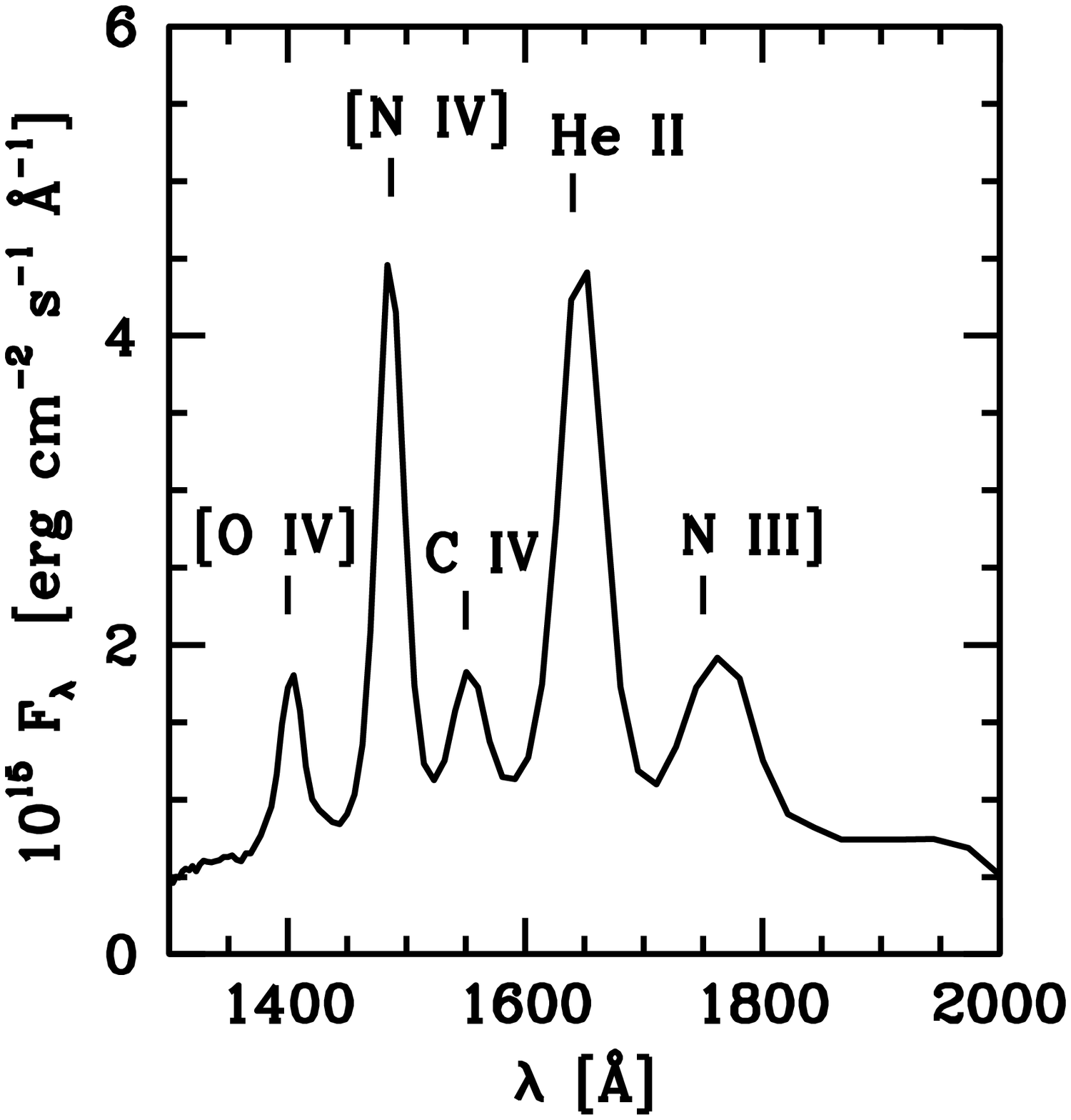}{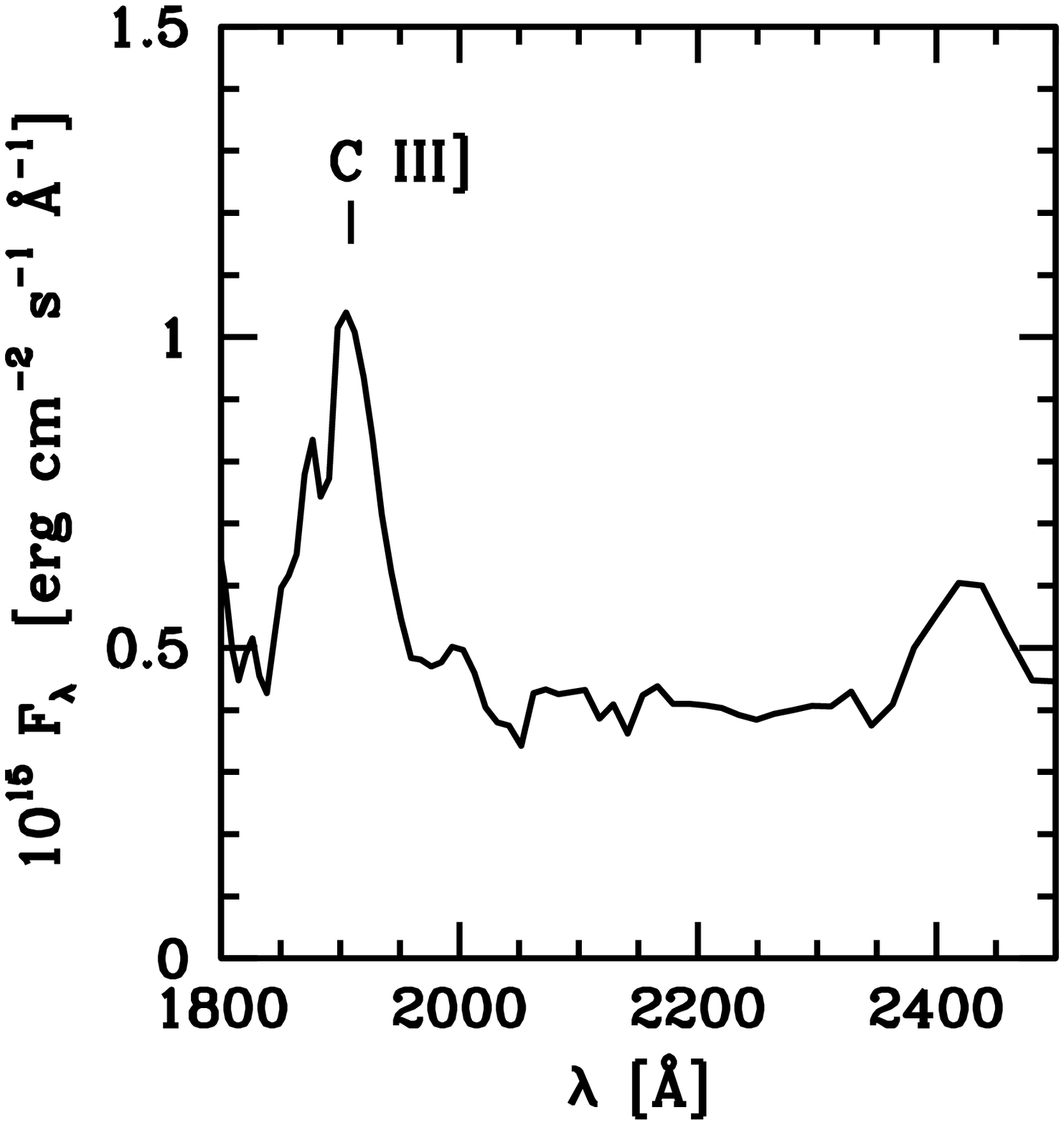}
\caption{As in Figure 1, for SMP~28.}
\label{fg:smp28}
\end{figure}

The tabular wavelength and flux of the final extracted spectrum was
then read using task {\tt tprint} and converted to 1-d spectrum using
the task {\tt rspectext}.  We averaged the multiple exposures using task {\tt
scombine}.  The final reduced spectra of the 11 PNe are shown in Figure
\ref{fg:smp6} to \ref{fg:smp28}.  The nebular line fluxes have been 
measured by integrating the area above the continuum using the IRAF task {\tt splot}
with the  {\tt d} or {\tt w} options.

In Table 2 we give the complete spectral analysis of our targets,  including the
identification wavelength (column 2), the line ID (column 3),
and the measured flux 
as a fraction of F$_{\rm H\beta}$ (column 4), where the H$\beta$ fluxes are from the {\it HST}
analysis of \citet{stanghellini03} and \citet{shawetal06}. It is worth noting that the emission around $\lambda$1650 could well be a blend of 
H II $\lambda$1640 and O III] $\lambda$1663 (the latter corresponding to the 1661/66 \AA~doublet). For those low- excitation PNe whose 
optical spectra do not show the $\lambda$4686 emission line (SMP~18, SMP~20, SMP~24) it is safe to assume that
the emission is due to O III]. In the other cases we identify the emission as a blend. We do not calculate 
abundances of other elements than carbon in this paper, so we do not present here a detailed analysis 
of the other emission lines.

Typical uncertainties in the emission line fluxes are $\sim5\%$ in the 
bright lines. 
We have examined data quality flags of each exposure, and noticed
that a few of the PR200L frames
have some saturated
pixels at the red end of their spectra.  This is caused by red pile-up, 
due to the low dispersion in the red wavelengths.  In general the saturated
pixels do not affect the measured fluxes. Only in two PNe, SMP~15 and SMP~18, the
saturated pixels affect the \ion{C}{2}]
$\lambda$2326 emission line, therefore these are listed as lower limits.

Furthermore, in some cases the spectra are affected by bad pixels. 
For SMP~24,
the 2D PR130L spectrum falls into an area of 
bad pixels, as indicated in the data quality file. If we exclude the 
bad pixels when extracting this spectrum we end up with a very
marred 1D spectrum, thus the \ion{C}{4} $\lambda$1550 line measurement is not reliable for this nebula,
even if  there seems to be a feature at the appropriate wavelength.
The PR200L spectrum looks much better, but the extension shows a dip
on the \ion{C}{3}] $\lambda$1909 line affecting the measured flux at an estimated level of 15$\%$ level.
For SMP~16 the bad pixels in the PR200L frame affect the measured fluxes of the 
\ion{C}{2}] $\lambda$2326 and \ion{C}{3}] $\lambda$1909 lines at an estimated level of 25$\%$.
Finally, in the case of SMP~18 the \ion{C}{2}] $\lambda$2326 
flux might be affected at the 15$\%$ level. In the case of SMP~25 a prominent NIV] $\lambda$1485 is
seen in the spectrum, but we do not give its flux in our tables since there are  few columns of bad pixels 
1425-1525\AA~ range for this PN.

Ultraviolet prism spectroscopy performed with the ACS have not been acquired often in the past, and
the analysis modes used in this paper are unique. In order to have a sanity
check of our calibration and spectral extraction we compare the ACS prism HRC/PR200L spectrum 
of  SMP~79, a PN in the LMC,  with the corresponding spectrum observed with STIS spectroscopy.
The prism spectrum was extracted in the same
way as for our SMC PN targets. We measured the flux of \ion{C}{3}]
$\lambda$1909 to compare with the flux measured from the STIS
spectrum.  In the {\it HST} archive there are a total of 24 prism exposures 
of SMP~79 taken at 12 different
positions of the CCD.  We extracted the
spectrum from each exposure, and measured the \ion{C}{3}] flux in each spectrum.  
The average flux is $(9.3 \pm 0.4) \times
10^{-13}$~ [erg~cm$^{-2}$~s$^{-1}$],
compared to the flux $9.4 \times 10^{-13}$
~[erg~cm$^{-2}$~s$^{-1}$] measured from the STIS
spectrum.  A contaminating source happened to lie just at
the spectral trace of \ion{C}{2}] $\lambda$2326, so we did not attempt
to compare this emission line with the one from STIS for LMC~SMP~79. 
Errors in our flux measurements are $\sim5\%$ 
except the specific cases listed above, as marked in Table 2.

\subsection{Extinction Correction}

The flux measurements need to be corrected both for Galactic
foreground extinction and for the SMC extinction proper.  
The relation between observed and
de-reddened fluxes, scaled to H$\beta$, can be written as 
\begin{equation}
 \frac{I_{\lambda}}{I_{\beta}} = \frac{F_{\lambda}}{F_{\beta}}10^{cf_{\lambda}},
\end{equation}
where $c$ is the target-dependent logarithmic extinction at H$\beta$
and $f_{\lambda}$ is the reddening function at wavelength $\lambda$.
Since the Galaxy and the SMC have different extinction curves in the UV
\citep{hutchings01}, we need to correct separately for each contribution.
We can write:
\begin{equation}
cf_{\lambda}=c_{\rm G}f_{\lambda, {\rm G}} + c_{\rm SMC}f_{\lambda, {\rm SMC}},
\end{equation}
where the suffixes "G" and "SMC" refer to the Galactic foreground and intrinsic SMC extinction, respectively.

\begin{figure}
\plotone{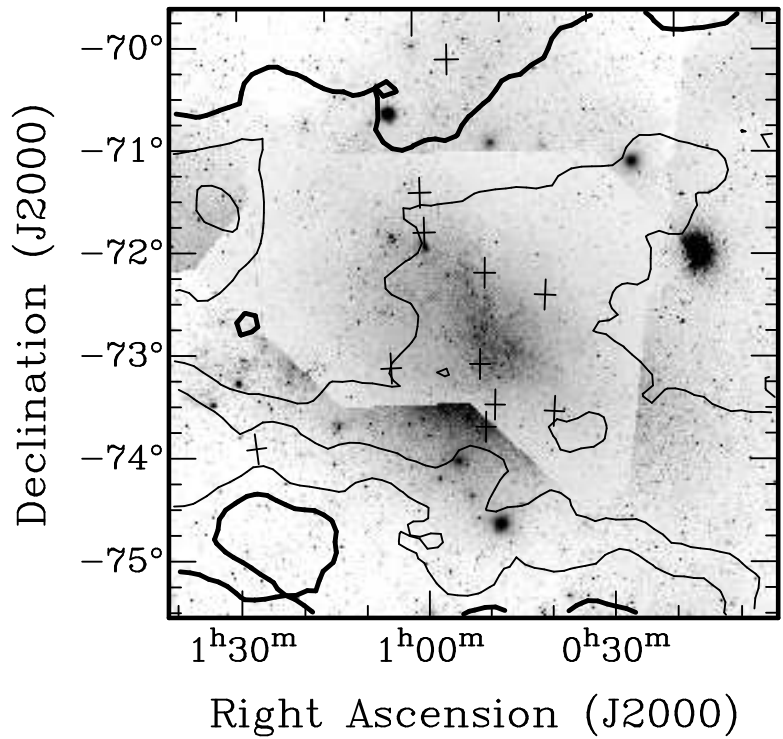}
\caption{The contours represent the Galactic foreground color excess map towards the SMC
are overlapped on the 
the Digital Sky Survey image of SMC.  The contours represent E$_{\rm B-V}$ between 0.04$^{\rm m}$
and 0.08$^{\rm m}$ from top to bottom, plotted for every 0.01$^{\rm m}$ interval, with the 
0.04 and 0.08 contours marked in thick lines. The positions of our targets are indicated with crosses.}
\label{fg:smc_ex}
\end{figure}

To evaluate the Galactic foreground extinction we used the Galactic
\ion{H}{1} foreground column density map constructed by
\cite{bot04} from data of the the Parkes \ion{H}{1} Survey of the
Magellanic System \citep{bruns05}.  The column density was
obtained by integrating the emission from $-60$ to $+50$ km s$^{-1}$,
outside the SMC velocities.  
The Galactic color excess map is calculated
using 
N(\ion{H}{1}) / E$_{\rm B-V}$ = 5.8 $\times 10^{21}$
[atoms~cm$^{-2}$~mag$^{-1}$] 
\citep{bohlin78, savage79}.  
Figure \ref{fg:smc_ex} shows the Galactic
foreground color excess map, superimposed to the Digital
Sky Survey\footnote{The Digitized Sky Surveys were produced at the
Space Telescope Science Institute under U.S. Government grant NAG
W-2166.  The images of these surveys are based on photographic data
obtained using the Oschin Schmidt Telescope on Palomar Mountain and
the UK Schmidt Telescope.  The plates were processed into the present
compressed digital form with the permission of these institutions.}
image of SMC. The target-specific foreground Galactic extinction constant,
$c_{\rm G} = 1.47 E_{\rm B-V}$, was estimated from Figure 13 for each target.

The constant $c_G$ is then used to correct the fluxes for
foreground Galactic extinction using the extinction function $f_{\lambda, {\rm G}}$ provided
by \cite{cardelli89}:
\begin{equation}
 \left(\frac{I_\lambda}{I_\beta}\right)_0=\frac{F_\lambda}{F_\beta}10^{c_{G}f_{\lambda, {\rm G}}}
\end{equation}

In order to correct for the SMC extinciton we estimate the optical extinction constant 
as:
\begin{equation}
 c_{\rm SMC}=2.875\log\frac{(H\alpha/H\beta)_0}{2.85},
\end{equation}
where $H\alpha/H\beta_0$ is the flux ratio corrected for
Galactic foreground.
We then correct the flux ratios for SMC extinction from:
\begin{equation}
 \frac{I_{\lambda}}{I_{\beta}}=\left(\frac{I_\lambda}{I_\beta}\right)_0 10^{{\rm c_{SMC}f_{\lambda, SMC}}}
\end{equation}
where the SMC the extinction function is from \cite{prevot84}.  
In Table 2 we list
the Galactic (column 5) and SMC (column 6) extinciton constants, as calculated above, 
and the final corrected intensity ratios (column 7). Note that the line extinction estimated in the foreground
overestimates the total line extinction in a few cases, thus the SMC extinction has been assumed to be zero.

\subsection{Comparison to Previous Observations}

While none of the observed targets had observations in the literature that allowed carbon abundance
determination, several have UV spectra taken in comparable wavelength ranges, and limited comparison could be
made among data sets. Let us examine these cases.

{\bf SMP~6}: The FOS spectrum published by \citet{vass96} is deemed to be inadequate for precise flux measurements
(see also SMP~28 below), thus we do not compare their fluxes to ours, although we do observe the same bright emission lines. 

{\bf SMP~15:} A low S/N IUE spectrum of this SMC PN, also known as N~43, has been acquired by \citet{aller87}. 
The only emission line available in the 1200-1910~\AA~ wavelength window is  CIII] $\lambda$1909, with
F$_{\lambda}=7\times 10^{-13}$ ~[erg cm$^{-2}$s$^{-1}$], which is within
$\sim$ 10 $\%$ of our measurement.

{\bf SMP~20:} Similarly to SMP~15, the only emission line measured by \citet{aller87} on the IUE spectrum of this
SMC PN, also known as N~54,
is C III] $\lambda$1909, with F$_{\lambda}=7\times 10^{-13}$~[erg cm$^{-2}$s$^{-1}$],  which is also
within $\sim$10$\%$ of our measurement.

{\bf SMP~28:} The UV spectrum of this SMC PN short-ward of the C III] $\lambda$1909 line is available from \citet{Mea90}.
\citet{Mea90} give flux values, and quote the flux error to better than 15$\%$ for brightest lines.
By comparing the flux of the bright emission lines that we have in common, 
O IV] $\lambda$1404, N IV] $\lambda$ 1487, He II $\lambda$1640, and N III $\lambda$1755, we found that they agree 
to better than the quoted errors for the two bluer lines, but the agreement 
gets worse for the redder lines measured in our PR130L spectrum. In addition to the IUE spectrum there is a FOS
spectrum of this PN \citep{vass96} whose fluxes are very uncertain, deemed to be systematically off by the authors. 
For this reason we will not compare our results with the FOS fluxes. We can use the FOS spectrum
as presented by \citet{vass96} to confirm that they also observe the C III] $\lambda$1909 emission line, contrary to what 
is derived from the IUE spectrum. Furthermore, the emission that we observe
at 1553 \AA~ is more likely to be associated with C IV than with Ne V, as suggested by \citet{Mea90}. In summary, the statement that this
PN is very carbon poor might not be correct in the light of our observations.

Finally, \citet{idiart} used the IUE final archived spectra to measure the $\lambda$1909 line for several of the PNe in our sample.
 The  $\lambda$1909 intensities by \citet{idiart} carry much larger
uncertainties than ours, due to the low signal-to-noise ratio of the IUE spectra. All intensities  agree with our data
within the uncertainties, except for SMP~6, SMP~13, and SMP~18 where the IUE spectra are noisy and the errors quoted by
\citet{idiart} are very large. The IUE spectra would not provide the CIV] nor the [C II] fluxes, thus their use to abundance
analysis can only establish  lower limits to the atomic abundances.

The comparison of the SMP~28 PR130L spectrum with that of \citet{Mea90} gives us confidence that the C IV] 
line flux that we measured is accurate. Overall, we do not have errors larger than 5-10$\%$ in this part of the spectrum. 
Furthermore, the comparison of the C III] emission lines of our PR200L observations with those by \citet{aller87}
confirms that we have reliable measurements for all the carbon lines.

\section{Abundance Analysis}

The ionic abundances of C$^{+}$/H$^+$ and C$^{2+}$/H$^+$ were computed using the {\tt
nebular} package in STSDAS \citep{shaw95, shaw98}. We used the line intensities of Table 2, and
T$_e$ and N$_e$ derived from the diagnostic lines measured by Shaw et al. (in preparation, hereafter SEA), if available, otherwise
we used the plasma diagnostics by \citet{leisy06}, hereafter LD06. In Table~3 we present the plasma diagnostics used in this paper; 
columns (2), (3), and (4) give respectively the low- and high-excitation T$_{\rm e}$ and the N$_{\rm e}$ used,
column (5) lists the references for the spectral lines used to calculate the diagnostics, and column (6) gives the 
excitation class (EC) of the nebulae, which have been estimated accordingly to Morgan (1984).

The C$^{3+}$/H$^+$ abundances must be derived from recombination lines, and we did so
on the basis of approximate relations from
\cite{aller84} (Eq. 5.40).  The ionic abundances for carbon are listed in Table 4,
where columns (2), (3), and (4) give respectively the C$^{+}$, C$^{2+}$ and C$^{3+}$ 
abundances in terms of hydrogen. 

It is worth noting that, at our spectral resolution, the $\lambda$2326 emission line could have a [O III] component from the
$\lambda$2321 emission. We calculate the volume emissivities for C$^+$ and O$^{2+}$, the latter from the optical emission lines, 
for all PNe where $\lambda$2326 has been observed. We found that the ratio of the emission volumes between the carbon and oxygen
transitions is typically $\sim$50 for the electron densities and temperatures of interest, thus the [O III] contribution does
not affect our C$^+$ ionic abundances. 

In order to calculate the total carbon abundances we follow the discussion by Kingsburgh \& Barlow (1994), also used by LD06.
For low excitation PNe where the C$^+$, C$^{2+}$, and C$^{3+}$  abundances are available, the total carbon abundance 
can be calculate simply by summing the ionic contribution. This is the case of SMP~13, SMP~15, SMP~16, SMP~18, and SMP~20, whose spectra do
not show He II emission thus correction for C$^{+4}$ would be unnecessary. This is also the case of SMP~6, whose 
emission around 1640-1663 \AA~ should be due almost entirely to the [O III]
component given the PN low optical excitation.

Next we need to calculate the ionization correction factors (ICFs) for the unseen ionization stages. We correct for the unseen C$^{+}$ lines
in SMP~8, SMP~24, SMP~25, SMP~26 and SMP~28 with Eqs. A11 and A13 (KB94). The ionic oxygen abundances used to obtain these 
ICFs have been calculated using the oxygen emission lines in SEA (SMP~8) and LD06,
the plasma diagnostics of Table 3, and the {\tt nebular} routines. As it turns out,
no correction is needed for SMP~8. In SMP~25, SMP~26, and SMP~28 we do not see the C$^{+}$ emission probably due to their high
excitation. We still calculate the correction and implement the correction, as in LD96, but this has a marginal effect on the final carbon 
abundances for these PNe.
We also calculated the ICFs
to account for the unseen C$^{4+}$ emission in the medium- to high-excitation PNe. For planetary nebulae SMP~25, SMP~26,
and SMP~28, showing He$^{2+}$ emission and not N$^{4+}$ lines, we use Eq. A20 (KB94). The ionic helium abundances needed in these
cases were calculated with the line intensities from LD06, the prescription of Benjamin et al. (1999), and the plasma diagnostics of Table 3.

The final ICFs are listed in column (5) of table 4, and the corrected abundances expressed in terms of A(C)=log(C/H)+12  are given in column (6),
together with a conservative estimate of their uncertainties. 

The larger factor of uncertainty for the total carbon abundances is the
uncertainty in the electron temperature. The errors listed in Table 4 include the propagation of
the uncertainty in the T$_{\rm e}$ for all ions, if the plasma diagnostics was available in SEA. For the other 
nebulae we estimate similar errors in the T$_{\rm e}$ and propagate the errors, as given in Table 4, 
unless there are other, larger uncertainties, as discussed below.

As reported in Table~2 the C$^{2+}$ fluxes of 
SMP~15 and SMP~18 are lower limits; we estimate that the total carbon abundances calculated
for these two PNe might be underestimated by $\sim5\%$.

Finally, while most of the emission lines intensities have errors $\sim$5$\%$, they go up to  
15 and 25$\%$ respectively in SMP~24 and SMP~16. These errors are propagated through
the carbon abundance analysis.

By considering the sample of this paper and the other carbon abundances in the literature, including only those who
consider all the present excitation levels (Aller 1987; LD06), we found that $<{\rm C/H}>_{\rm SMC}$=(3.75$\pm$3.64)$\times$10$^{-4}$.
This is $\sim$1.5 times higher than the same average for the LMC ($<{\rm C/H}>_{\rm LMC}$=(2.49$\pm$2.18)$\times$10$^{-4}$, Stanghellini et al.
2005).

\section{Magellanic Cloud PNe and the stellar evolution models}

Carbon, the fourth most abundant element in the Universe \citep{clayton03}, is vigorously produced in LIMS, 
thus it probes their evolution.
The $\alpha$-elements (oxygen, neon, argon and sulfur), on the other hand,  are produced by nucleosynthesis
of Type II supernova and provide information about the original composition
of the PN progenitor at the time of birth. Oxygen may be brought up to the LIMS surface by the 
third dredge-up, and its abundance should be used cautiously in this capacity.

We use the SMC PN carbon abundances determined in this paper, and those available in the literature, and relate them
to the abundance of the $\alpha$-elements, to assess the models of stellar evolution and the 
theoretical yields. In order to compare populations of different metallicity we also include in the plots the results from our study of
carbon abundances in the LMC PNe \citep{stanghellini05}. 
The abundances of N, O, and Ne
for LMC PNe used here are from \citet{aller87}, and Leisy \& Dennefeld (1996, 2006). 
Corresponding values of N, O, and Ne in SMC PNe come from \citet{leisy96}, and SEA. 

\begin{figure}
\plotone{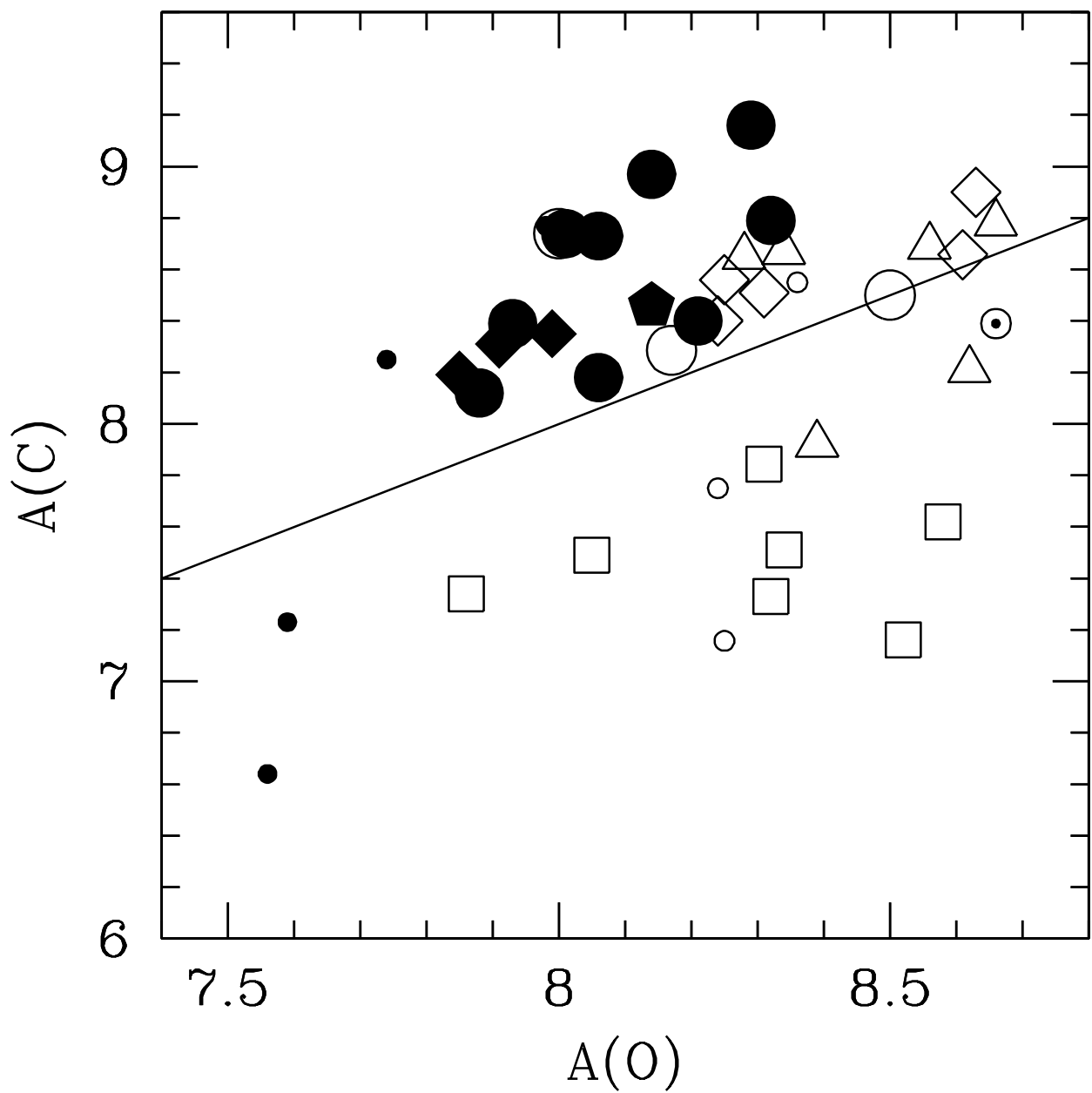}
\caption{The distribution of SMC (filled symbols) and LMC (open symbols) PNe on the log(C/H) -- log (O/H) plane. Symbol shapes represent the different morphologies. Circles: Round, diamonds: Elliptical, triangles: Bipolar core, squares: Bipolar, pentagons: Point-symmetric PNe. Small circles refer to unknown morphology, see text. The thick solid line represents the 1:1 correlation. 
The solar symbol indicates the abundances from Asplund et al. (2005).}
\label{fg:ch_oh}
\end{figure}

In Fig. \ref{fg:ch_oh}  we show the carbon vs.~oxygen abundances\footnote{For the sake of clarity we do not plot carbon abundance error bars, that are given in Table 4. Note that uncertainties of $\sim$10$\%$ would be  within the symbol size.}  of PNe in the SMC (filled symbols) and the LMC (open symbols).  Their morphologies are indicated by symbols of various shapes.  Small symbols are used for PNe of unknown morphologies, i.~e., not yet 
observed with  {\it HST} or, in a couple of cases, where the morphological class was too uncertain to be assigned. Abundances are in the usual scale of A(X)=log(X/H)+12. The PNe separate into two groups: The first group are those with C/O$<$1 (below the line), whose shapes are almost exclusively bipolar or bipolar core (BC\footnote{PNe with BC morphologies are those with pinched or barrel-shaped cores similar to known bipolars with protruding lobes.  This classification is somewhat ambiguous and may include some PNe with unusual shapes, and some misclassified elliptical PNe.}), as already noted in Stanghellini et al. (2007); these PNe are associated with relatively young stellar populations, such as the disk of the Milky Way. And the second group are those with C/O$>$1, corresponding to the less massive progenitors. 
Most  of the SMC PNe lie in the C/O$>$1~ part of this plot.

The predominance of C-rich PNe in the SMC is consistent with the theoretical
expectation that the third dredge-up is predicted to be favored at
lower metallicity, i.~e., to occur in stars of lower masses and with a higher efficiency
(e.~g., Karakas, Lattanzio, \& Pols 2002; 
Stancliffe et al. 2005).  Carbon is converted to nitrogen by hot-bottom burning (HBB).  HBB is likely active only in relatively high-mass AGB stars (M$^{\rm min}_{\rm HBB} \sim$ 3.5--5 M$_{\odot}$, depending on metallicity ) with very deep convection and hot cores (Marigo 2001).  Star formation has been far more active in the gas-rich LMC than in the gas-depleted SMC.  Possibly, many of the PNe that have formed in the LMC have higher average masses than those in the SMC, affecting the fraction of high-end mass LIMS that will go through the HBB phase.  That is, the segregation of PNe with bipolar and BC symmetries below the C=O line is fully consistent with model calculations. It is worth recalling, in fact, that
Villaver (2004) found a lack of intermediate-mass central stars in the SMC, present in the LMC.

SMC SMP~22 and SMC SMP~25 are two exceptions to this scenario.  They are the two morphologically unclassified SMC PNe in the lower-left of the plot: low metallicity (LD06) and C/O $< $1. Not much is known about SMP~22.  SMP~25 is quite unique among SMC PNe. It is the only PN in the SMC where oxygen-rich dust has been detected (Stanghellini et al. 2007), while all other SMC PNe observed with Spitzer/IRS have carbon-rich dust. Furthermore, the central star of SMP~25 is much more massive (M$\sim$0.82 M$_{\odot}$, Villaver 
et al. 2004) than the typical central in Magellanic Cloud PNe (Villaver et al. 2007), and it is  located in the eastern region of the SMC facing the LMC, a region that contains younger and more metal rich clusters \citep{crowl01} and where \citet{dopita85} found the kinematically younger
PNe to be concentrated. SMP~25 may have undergone recent HBB activity. 

\begin{figure}
\plotone{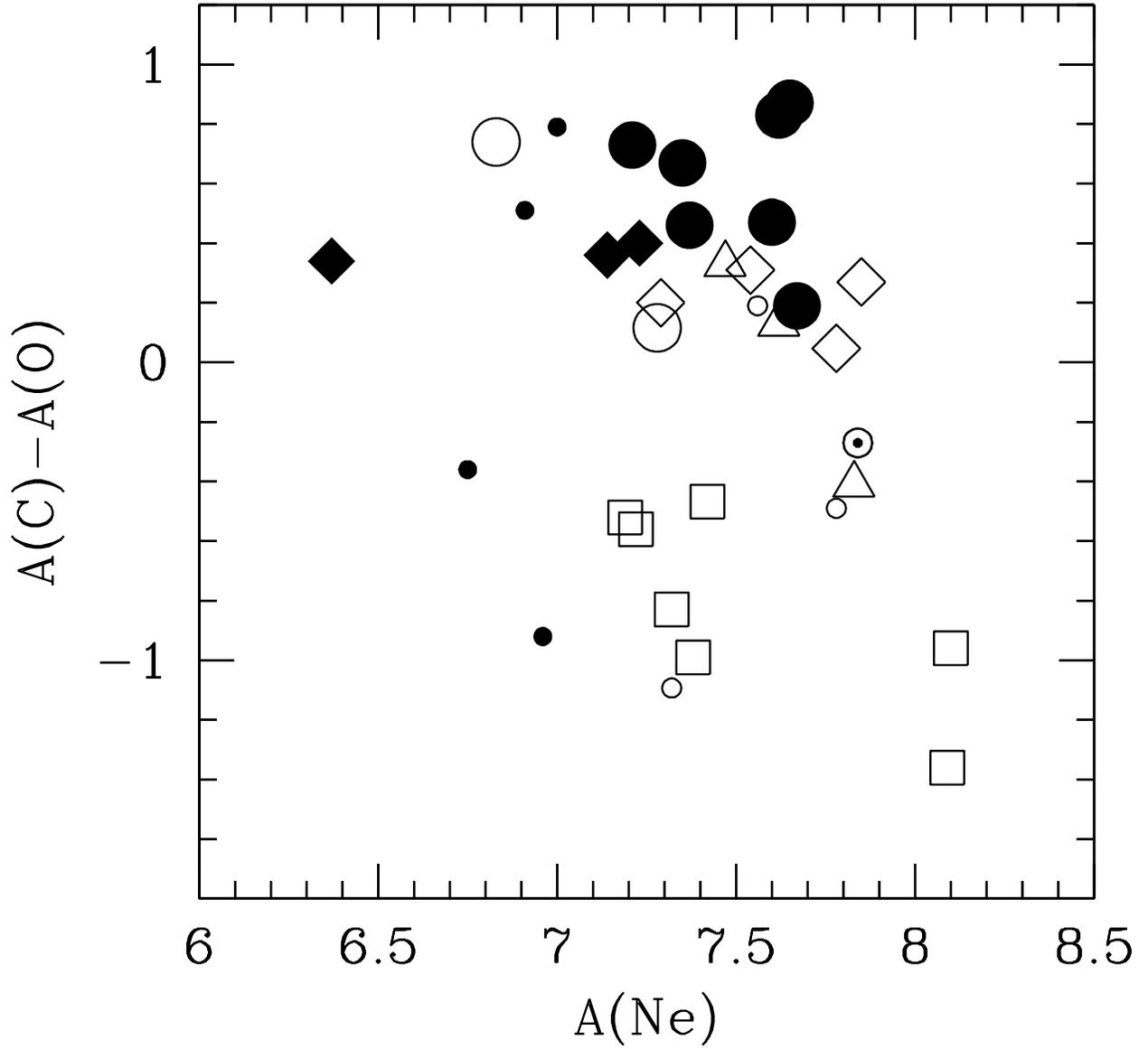}
\caption{The distribution of SMC and LMC PNe on the log(C/O) -- log (Ne/H) plane. Symbols  are used as in Fig. \ref{fg:ch_oh}}
\label{fg:co_ne}
\end{figure}

In Figure \ref{fg:co_ne}, where we show the relation between C/O and the neon abundance, we confirm the carbon-abundance segregation of SMC PNe. There is no evidence that neon's abundance has been modified in any way, nor that the C/O ratio is modified differently 
in the SMC than in the LMC depending on the initial neon.

\begin{figure}
\plotone{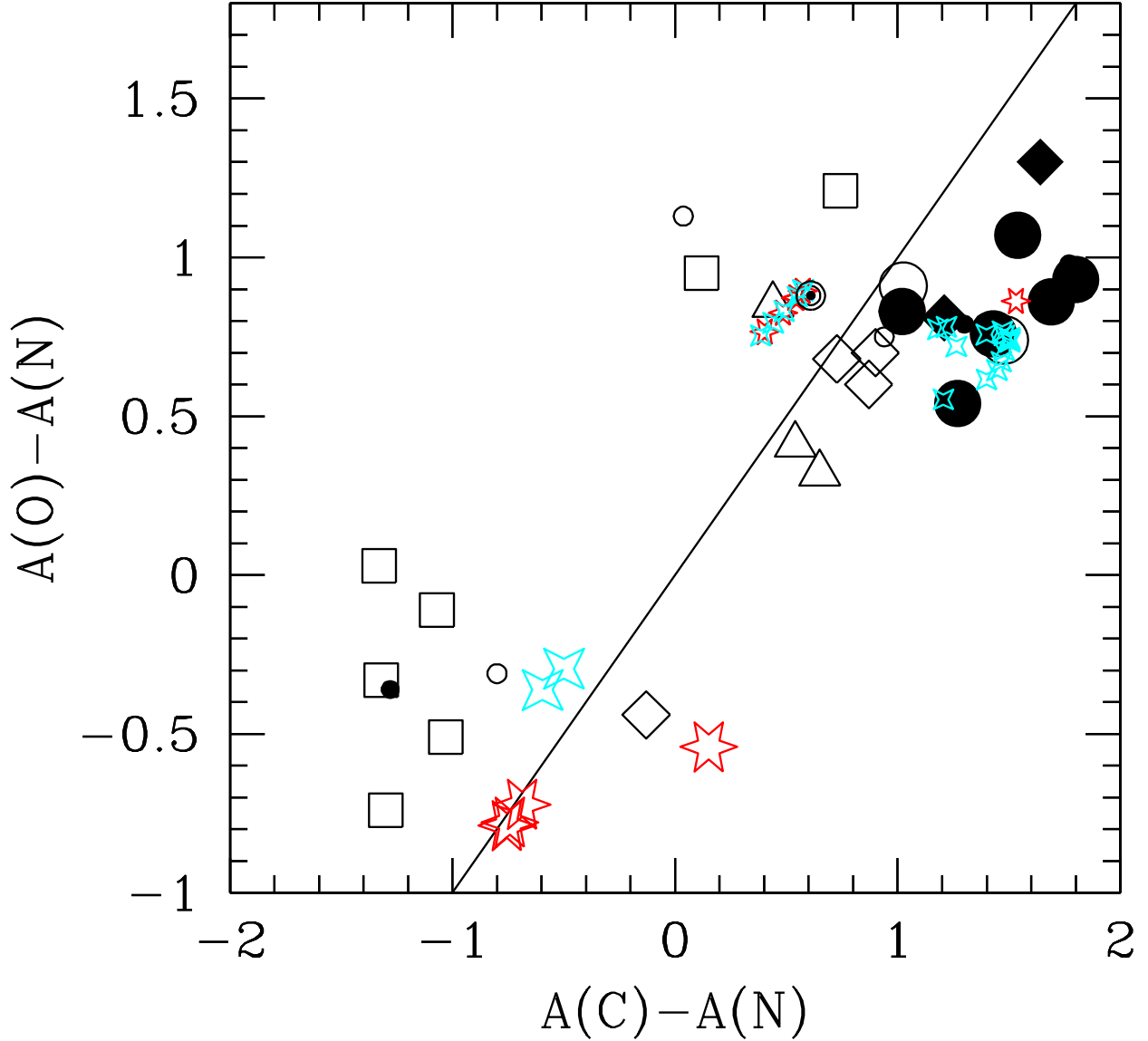}
\caption{The distribution of SMC and LMC PNe on the log(C/N) -- log(O/N) plane. Symbols for data are used as in Fig. \ref{fg:ch_oh}. The starred symbols represent the yields from stellar evolution by \citet{marigo01}, where the four-points stars are for the LMC and six-points stars are for the SMC PNe. Within the same set of models, different points indicate different initial mass, and the smaller symbols represent the yields from the evolution of M$_{\rm to}<$M$^{\rm min}_{\rm HBB}$. The sun locus would be at (0.61; 0.88) in this plot,
but it is not marked to avoid overlap with a target and few model points.}
\label{fg:on_cn}
\end{figure}

The  log(O/N) vs.~log(C/N) plot (Fig.~\ref{fg:on_cn}) also shows some obvious and interesting groupings of data. The segregation of bipolar
nebulae on the left side of the plotted line is expected of course, since figures 13 and 15  are not entirely independent.  However, it is clear that the bipolar that exhibit O $>$ C also have N$>$C: that is, they are "N rich".   Symbiotic stars and novae stars (not shown;  Nussbaumer et al. 1988) are also found in the same region.  On the other hand, all of the sample with C/O $>$ 1 from Fig. 13 are also nitrogen poor; that is, N $<$ O and N$ <$ C.  Most of the SMC PNe fall into the latter group, of course.  Carbon stars (not shown;  Nussbaumer et al. 1988) are also found in this region.  

\begin{figure}
\plotone{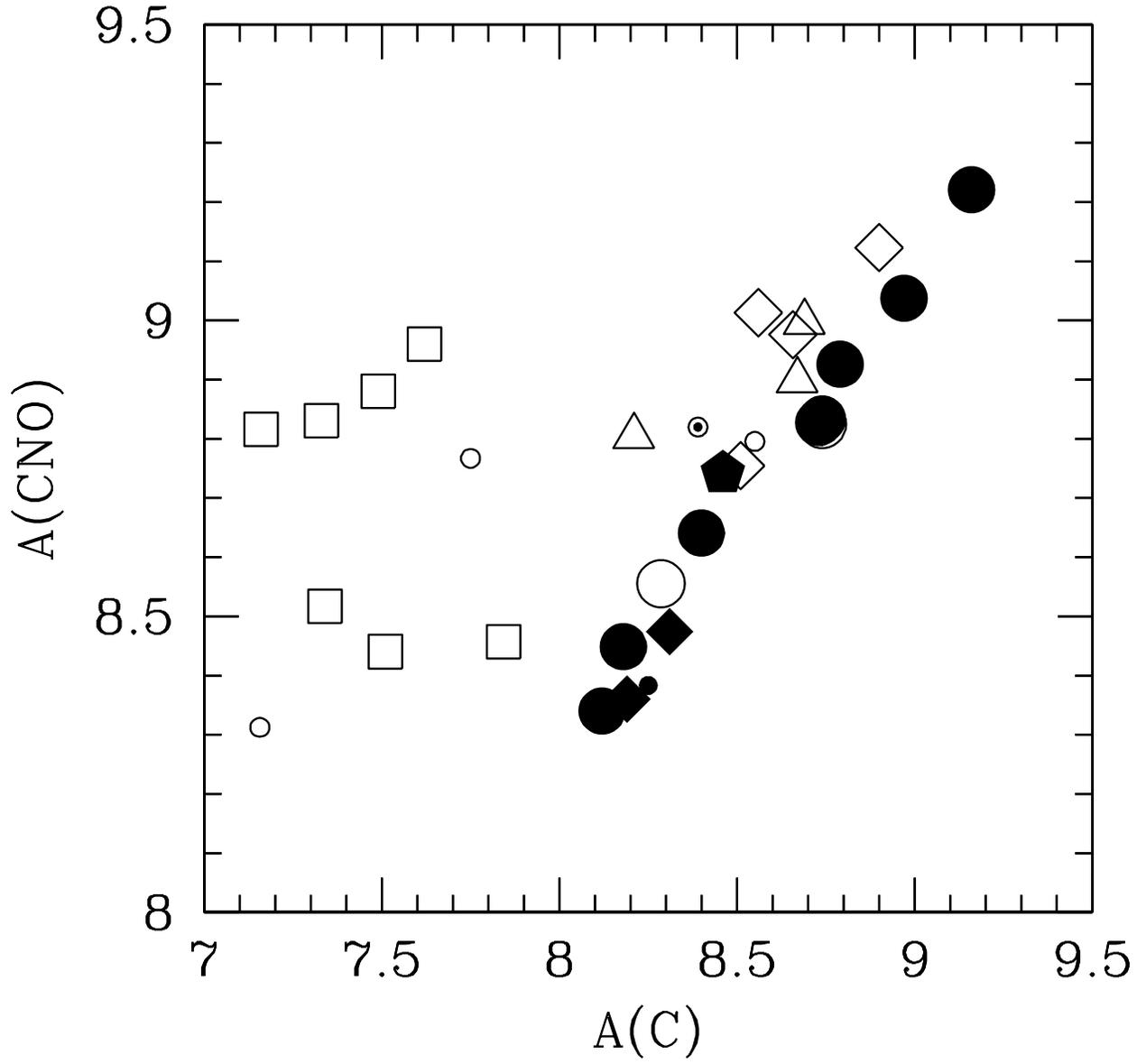}
\caption{SMC and LMC PN abundance sum of C, N, and O vs. carbon abundance. Symbols are as in Fig. \ref{fg:ch_oh}.}
\label{fg:cno_c}
\end{figure}

Carbon stars probably have not undergone any HBB processing.  
We can compare the data in Fig.~15 with the yields form models of stellar evolution. We limit our comparison to the final yields calculated by \citet{marigo01}, to avoid overcrowding of the plot. The yields by \cite{vdhg} or \citet{karakas07} would look very similar to the ones by \citet{marigo01} on this plot.
We indicate the yields from LIMS evolution with starred symbols: the four-pointed stars correspond Z=0.008 while the six-pointed stars are for Z=0.004, and we used smaller symbols for stars with progenitor masses too low to have undergone the HBB (M$^{\rm min}_{\rm HBB} \sim$  3.5 and 4.0 M$_{\odot}$ for the SMC, and 
the LMC models respectively, as in Marigo 2001). The abundance ratios measured for round and elliptical PNe correspond very 
well with the final yields predicted for the evolution of the lower mass-end of LIMS in both the SMC and the LMC.  Similarly, the {\it massive} models are in good agreement with the abundance ratios of C, N, and O in the PNe in the LMC with bipolar and BC morphologies.  These results are extremely satisfying verifications of difficult model predictions.
In Figure \ref{fg:cno_c} we plot the sum of carbon, oxygen, and nitrogen abundances vs.~carbon of LMC and SMC PNe.  Fig.\ref{fg:cno_c} is a classical diagram for assessing the efficiency of CNO-cycling . The CNO abundance would be preserved if these elements were only acting as catalysts, while carbon enhancement via
third dredge up is clearly present in the SMC PNe. 

\begin{figure}
\plotone{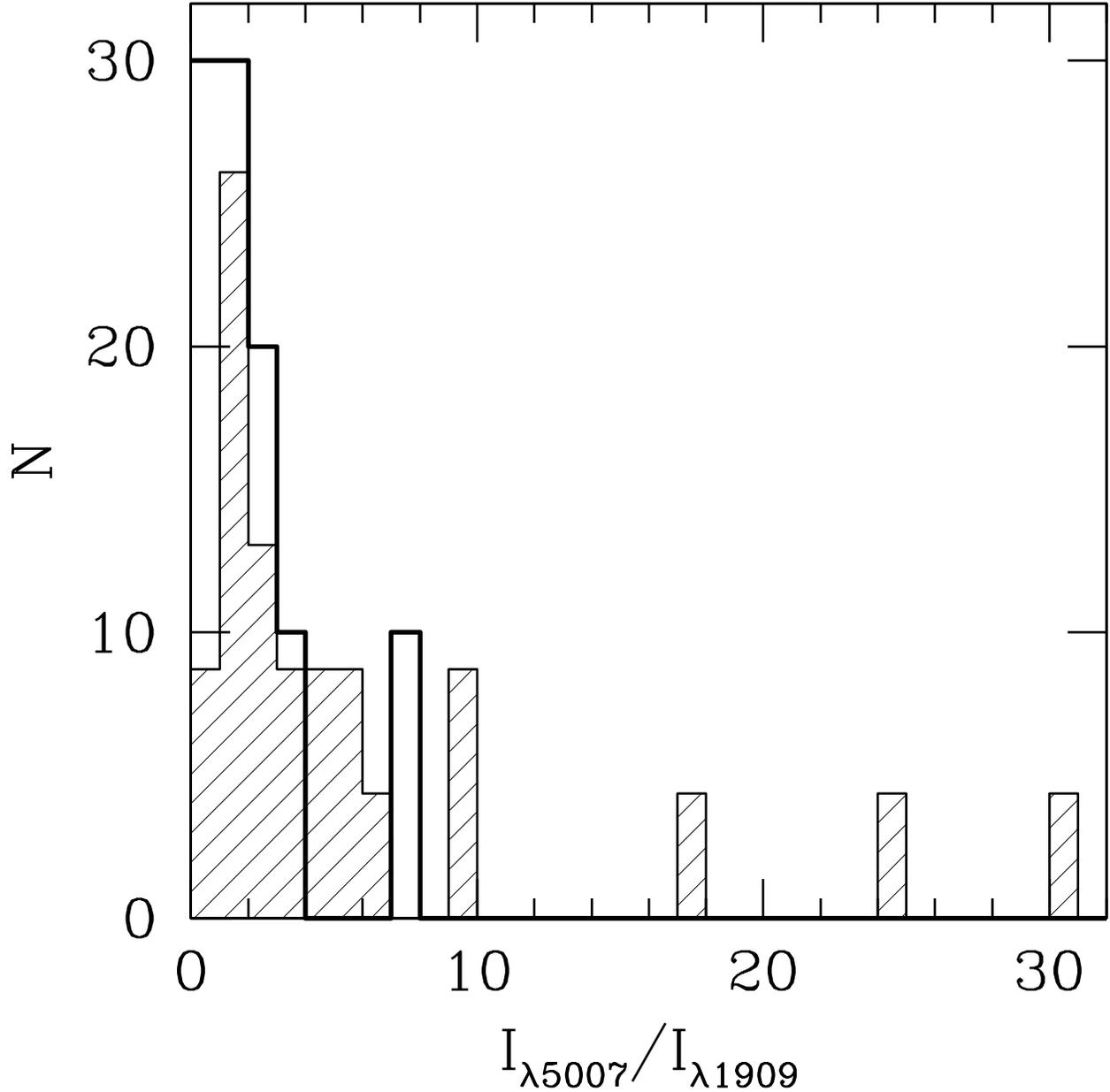}
\caption{Cooling of SMC and LMC PNe through the major emission lines. The I$_{\lambda5007}$/I$_{\lambda1909}$ histogram is shown
for the SMC (heavy line) and the LMC (light shaded histogram) PNe. Note the very different distributions for I$_{\lambda5007}$/I$_{\lambda1909}<$ 1 and $>$6.}
\label{fg:histo}
\end{figure}

Based on a study of LMC and SMC optical lines,  \citet{stanghellini03}  found that the 
observed intensity ratio ${\rm I_{[O III] \lambda5007} / I_{H\beta}}$ seems to increase with metallicity. The photoionization 
models in \citet{stanghellini03} show that metallicity impacts the relative emission line 
strength of the major coolants, and that the UV carbon emission lines have a significant
effect in cooling PNe with low metallicities.  Now that we have observations of carbon emission lines in both SMC and LMC PNe we can assess the validity of those models. In  Fig.\ref{fg:histo} we plot the histogram of the I$_{\lambda5007}$/I$_{\lambda1909}$ line intensity ratio
for the SMC (thick line) and the LMC (thin, shaded histogram) PNe. The distributions, which  have been normalized for the number of PNe in each sample, appear to be different at very low and very high intensity ratios, indicating that the carbon emission line is a very important coolant for most  SMC PNe. We calculate that the average line intensity ratio is 6.23 and 2.36 respectively for the LMC and SMC PNe,
and the median values of I$_{\lambda5007}$/I$_{\lambda1909}$  are  3.34$\pm$2.98 and 1.83$\pm$0.83 for the LMC and the SMC PNe respectively, marking a sharp difference in the dominant cooling agents at different metallicities. The carbon $\lambda$1909 line is a much more efficient coolant, with respect to the $\lambda$5007 line, in the SMC than the LMC PNe,
The CIV line can also be an important coolant in the low-metallicity, highly ionized  PNe.

\section{Conclusions}

The present paper provides a database of 11 SMC PNe whose carbon abundance has been accurately derived through prism spectroscopy with the ACS/{\it HST}. The number of reliable SMC PNe with well-measured carbon abundances has doubled, 
finally allowing statistically useful analysis of carbon production and CNO-cycling in PNe at the very low metallicity in the progenitors of SMC PNe.  

By comparing the abundances of SMC and LMC PNe we found that most observed SMC PNe are carbon rich, except in a couple of unusual cases, SMP~22 and SMP~25, where there is indication that the progenitors underwent the HBB process. This seems to indicate that most SMC PNe derive from low-mass (M$_{\rm to}<$3.5 M$_{\odot}$) and low-metallicity progenitors. By comparison, the LMC has a much varied PN population, where both carbon-rich and carbon-depleted PNe are present, indicating a larger range of PN progenitor mass and metallicity, including several PNe whose progenitors could be in the 3-5 M$_{\odot}$ mass range. While a larger SMC PN sample would improve the impact of these 
findings, it is worth recalling that both LMC and SMC samples were selected homogeneously, and that with the present observations we have increased the statistical significance of the SMC PN sample to almost the same level of confidence as the LMC sample (Shaw et al. 2006).

The CNO abundances in Magellanic Cloud PNe can now be used to test the predictions of models 
of stellar evolution. We found that the data agree impressively with the 
final
yields from Marigo (2001) models, calculated both for the SMC and LMC metallicities. In particular, the yields calculated for turnoff mass 
$<$3.5, and 4.0, M$_{\odot} $~seem to reproduce PN abundances of round and elliptical PNe in the SMC, and in the LMC, respectively. Yields
from the more massive LIMS encompass well the bipolar PNe abundances. If we accept the model results then the production of carbon through the CNO cycle in the SMC PNe shows that most SMC PN progenitors are in the low-mass end of the LIMS mass range, while the LMC PNe could have progenitors in the whole LIMS range. In some cases extra-mixing or other process that were not included in the current models are needed to justify the observed CNO abundances.

Compared to the SMC, the LMC has a been a site of recent or ongoing star formation and heavy-element enrichment.  One of the most important results of our study is that the average abundance of carbon in SMC PNe is $\sim$ 1.5 times higher than the average carbon measured in LMC PNe. Using the Magellanic Cloud sample of PNe with determined carbon abundances, and whose IRS/Spitzer spectra allows a determination of the dust chemistry (Stanghellini et al. 2007) we find that $<$C/H$>$=(4.69 $\pm$3.33)$\times$10$^{-4}$ for Magellanic Cloud PNe with carbon-rich dust.  This is $\sim$35 times  greater than the same ratio for PN with oxygen-rich dust, $\ <$C/H$>$=(1.35 $\pm$0.88)$\times$10$^{-5}$, confirming a strong correlation between dust and gas chemistry.

The present observations also show that the high-excitation carbon emission lines are major or dominant coolants in the low metallicity SMC PNe, as predicted by Stanghellini et al. (2003), much more so than in LMC PNe, where the C III] $\lambda$1909 line intensity is typically a small fraction 
of the [O III] $\lambda$5007 line strength.

\section{Acknowledgements}

We thank Mark Dickinson for his help in modeling the prism response during the ACS  phase II. Thanks are due to an anonymous Referee for important suggestions.
Support for this work was provided by NASA through grant GO-10250.01-A from Space Telescope 
Science Institute, which is operated by the Association of Universities for Research
in Astronomy, Inc., under NASA contract NAS5-26555.

\clearpage 

\clearpage

\clearpage

\clearpage

\clearpage 

\clearpage 

\begin{deluxetable}{clllr}
\tablecaption{Observing logs\label{tb:log}}
\tablehead{ 
\colhead{Name} & \colhead{Data Set} & \colhead{Observation Date} &
\colhead{Setup} & \colhead{t$_{exp}$} \\
& & & & (s)
}
\startdata
SMP~6 & j90d01011 & 2005 May 03 & HRC/PR200L & 2656 \\
& j90d01f7q & 2005 May 03& HRC/F330W & 360 \\
& j90da1fbq & 2005 May 03& SBC/PR130L & 2825 \\ 
& j90da1faq & 2005 May 03& SBC/F165LP & 360 \\
SMP~8 & j90d02011 & 2006 Jan 18 & HRC/PR200L & 2716 \\
& j90d02kjq & 2006 Jan 18& HRC/F330W & 300 \\
& j90da2kmq & 2006 Jan 19 & SBC/PR130L & 2820 \\
& j90da2knq & 2006 Jan 19& SBC/F165LP & 360 \\
SMP~13 & j90d03021 &  2005 Aug 05 & HRC/PR200L & 2656 \\
& j90d03011 & 2005 Aug 05& HRC/F330W & 306 \\
& j90da3y4q & 2005 Aug 05& SBC/PR130L & 2820 \\
& j90da3y5q & 2005 Aug 5& SBC/F165LP & 360 \\
SMP~15 & j90d04011 & 2005 May 03 & HRC/PR200L & 2656 \\
& j90d04fiq & 2005 May 03& HRC/F330W & 360 \\
& j90da4fmq & 2005 May 03& SBC/PR130L & 2820 \\
& j90da4flq & 2005 May 03& SBC/F165LP & 360 \\
SMP~16 & j90d05011 & 2005 Oct 01 & HRC/PR200L & 2656 \\
& j90d05saq & 2005 Oct 01& HRC/F330W & 360 \\
& j90da5sdq & 2005 Oct 01& SBC/PR130L & 2829 \\
& j90da5seq & 2005 Oct 01& SBC/F165LP & 360 \\
SMP~18 & j90d06011 & 2005 Feb 14 & HRC/PR200L & 2656 \\
& j90d06drq & 2005 Feb 14& HRC/F330W & 360 \\
& j90d16p2q & 2006 Jan 12 & SBC/PR130L & 2640 \\
& j90d16p9q & 2006 Jan 12& SBC/F165LP & 300 \\
SMP~20 & j90d07011 & 2005 Apr 22 & HRC/PR200L & 2656 \\
& j90d07aoq & 2005 Apr 22& HRC/F330W & 360 \\
& j90da7arq & 2005 Apr 22& SBC/PR130L & 2820 \\
& j90da7asq & 2005 Apr 22& SBC/F165LP & 360 \\
SMP~24 & j90d10011 & 2006 Sep 03 & HRC/PR200L & 2656 \\
& j90d10aqq & 2006 Sep 03& HRC/F330W & 360 \\
& j90da0atq & 2006 Sep 03 & SBC/PR130L & 2825 \\
& j90da9auq & 2006 Sep 03& SBC/F165LP & 360 \\
SMP~25 & j90d08u5q & 2006 Sep 01 & HRC/PR200L & 2702 \\
& j90d08u6q & 2006 Sep 01& HRC/PR200L & 3306 \\
& j90d09u4q & 2006 Sep 01& HRC/F330W & 360 \\
& j90da8010 & 2006 Sep 01 & SBC/PR130L & 6142 \\
& j90da8u9q & 2006 Sep 01& SBC/F165LP & 360 \\
SMP~26 & j90d09s4q & 2005 Sep 30 & HRC/PR200L & 2702 \\
& j90d09s5q & 2005 Sep 30& HRC/PR200L & 3305 \\
& j90d09s3q & 2005 Sep 30& HRC/F330W & 360 \\
& j90da9010 & 2005 Oct 01 & SBC/PR130L & 6137 \\
& j90da9s8q & 2005 Oct 01& SBC/F165LP & 360 \\
SMP~28 & j90d11011 & 2005 Aug 02 & HRC/PR200L & 2656 \\
& j90d11gsq & 2005 Aug 02& HRC/F330W & 360 \\
& j90db1gvq & 2005 Aug 02& SBC/PR130L & 2820 \\
& j90db1gwq & 2005 Aug 02& SBC/F165LP & 360 \\
\enddata
\end{deluxetable}

\begin{deluxetable}{lrlrrrr}
\tablecaption{Spectral analysis} 

\tablehead{ 

\colhead{Name} &

\colhead{$\lambda$} & 

\colhead{ID} & 

\colhead{F$_{\lambda}$\tablenotemark{a}} &

\colhead{c$_{\rm G}$}&

\colhead{c$_{\rm SMC}$}&

\colhead{I$_{\lambda}$\tablenotemark{a}}\\
}

\startdata 

     SMP~06  & 1335  &     \ion{C}{2}]   &       2.73  & 0.078  & 0.320  &      42.37 \\
     	     & 1550  &      \ion{C}{4}  &      20.25  & 0.078  & 0.320  &     156.10 \\
             & 1640-1663  &     \ion{He}{2},\ion{O}{3}]  &       9.15  & 0.078  & 0.320  &      60.62 \\
             & 1909  &    \ion{C}{3}]  &     184.20  & 0.078  & 0.320  &     852.60 \\
             & 2326  &     \ion{C}{2}]  &      19.12  & 0.078  & 0.320  &      57.88 \\
    SMP~08  & 1550  &      \ion{C}{4}\tablenotemark{b}  &      16.98  & 0.026  & 0.00  &      18.29 \\
             & 1909  &    \ion{C}{3}]  &     298.90  & 0.026  & 0.00  &   323.32   \\     
     SMP~13  & 1550  &      \ion{C}{4}  &      40.46  & 0.076  & 0.130  &     104.80 \\
     	     & 1909  &    \ion{C}{3}]  &     501.90  & 0.076  & 0.130  &    1062.00 \\
             & 2326  &     \ion{C}{2}]  &      34.82  & 0.076  & 0.130  &      62.84 \\

     SMP~15  & 1550  &      \ion{C}{4}\tablenotemark{b}  &       2.45  & 0.01  & 0.00  &   2.52  \\
             & 1909  &    \ion{C}{3}]  &     221.00  & 0.01  & 0.00  &   227.8  \\
             & 2326  &     \ion{C}{2}]  &      $\ge$31.85  & 0.01  & 0.00  &   $\ge$32.82 \\
    SMP~16  & 1550  &      \ion{C}{4}\tablenotemark{b}  &       6.43  & 0.03  & 0.00  &  7.01     \\
             & 1909  &    \ion{C}{3}]  &      89.03\tablenotemark{c}  & 0.03  & 0.00  &   97.47    \\ 
             & 2326  &     \ion{C}{2}]  &      93.97\tablenotemark{c}  & 0.03  & 0.00  &    102.82 \\ 

     SMP~18  & 1550  &    \ion{C}{4}  &       0.68  & 0.076  & 0.058  &       1.17 \\
       & 1663  &     \ion{O}{3}  &       2.96  & 0.076  & 0.058  &       4.90 \\
       & 1909  &    \ion{C}{3}]  &     113.80  & 0.076  & 0.058  &     179.20 \\
       & 2326  &     \ion{C}{2}]  &      $\ge$62.16  & 0.076  & 0.058  &      $\ge$92.37 \\ 

    SMP~20  & 1550  &      \ion{C}{4}\tablenotemark{b}  &       6.88  & 0.00  & 0.00  &    6.88\\ 
       & 1663  &     \ion{O}{3}]  &       5.93  &   0.00&  0.00 &       5.93 \\
       & 1909  &    \ion{C}{3}]  &     528.30  &  0.00& 0.00  &     528.3 \\
       & 2326  &     \ion{C}{2}]  &      87.06  &  0.00& 0.00  &      87.06 \\
    SMP~24  & 1663  &     \ion{O}{3}]  &       2.18  & 0.047  & 0.00  &     2.49   \\
       & 1909  &    \ion{C}{3}]  &     147.60\tablenotemark{d}  & 0.047  & 0.00  &     168.43 \\ 
     SMP~25  & 1397-1407  &   [\ion{O}{4}]  &      65.93  & 0.069  & 0.042  &     109.10\\ 
       & 1550  &      \ion{C}{4}  &      68.02  & 0.069  & 0.042  &     104.40 \\
       & 1640-1663  &     \ion{He}{2} ,\ion{O}{3}]  &     337.30  & 0.069  & 0.042  &     504.40 \\
       & 1750  &     \ion{N}{3}]  &     106.50  & 0.069  & 0.042  &     155.10 \\
       & 1909  &    \ion{C}{3}]  &      31.06  & 0.069  & 0.042  &      44.87 \\
     SMP~26  & 1550  &      \ion{C}{4}  &     509.50  & 0.075  & 0.190  &    1853.00\\ 
       & 1640- 1663  &     \ion{He}{2}, \ion{O}{3}] &     158.20  & 0.075  & 0.190  &     524.10 \\
      & 1909  &    \ion{C}{3}]  &     703.40  & 0.075  & 0.190  &    1895.00 \\
     SMP~28  & 1397-1407  &     [\ion{O}{4}]  &      36.48  & 0.098  & 0.041  &      65.49\\ 
       & 1487  &     [\ion{N}{4}]  &     162.00  & 0.098  & 0.041  &     275.40 \\
       & 1550  &      \ion{C}{4}  &      61.75  & 0.098  & 0.041  &     102.20 \\
       & 1640-1663  &     \ion{He}{2} ,\ion{O}{3}  &     257.30  & 0.098  & 0.041  &     413.50 \\
       & 1750  &     \ion{N}{3}]  &      88.24  & 0.098  & 0.041  &     138.00 \\
      & 1909  &    \ion{C}{3}]  &      50.10  & 0.098  & 0.041  &      78.34 \\

\enddata
\tablenotetext{a}{In terms of H$\beta$=100, where optical fluxes are from \citet{stanghellini03}, \citet{shaw06}.}
\tablenotetext{b}{P-Cygni absorbed component was also detected corresponding to this feature. The
flux corresponds to the redshifted emission line.}
\tablenotetext{c}{Flux uncertainty$\sim25\%$}
\tablenotetext{d}{Flux unceratinty$\sim15\%$}

\end{deluxetable}

\begin{deluxetable}{lrrrlr}
\tablecaption{Plasma Diagnostics} 
\tablehead{
\colhead{Name} & 
\colhead{T$_{\rm e}$\tablenotemark{a}} &
 \colhead{T$_{\rm e}$\tablenotemark{b}} &
\colhead{N$_{\rm e}$\tablenotemark{b}} & 
\colhead{ref}&
\colhead{EC}\\

\colhead{} & 
\colhead{[K]} & 
\colhead{[K]} & 
\colhead{[cm$^{-3}$]} & 
\colhead{} &
\colhead{}\\}
\startdata

        SMP~6     &   15600     &   14400  &  11400  &    LD06& 4\\
    
        SMP~8     &     \dots       &  13700$^{+1110}_{-1170}$  &    2770  &  SEA& 2-3\\ 
    
       SMP~13     &    \dots       &  12800$^{+1080}_{-940}$  &  2900  &  SEA& 4\\  
           
       SMP~15     &    16200    &  12000  &   5000:&     LD06& 2-4\\ 
    
       SMP~16     &  \dots  &   11800  &   5000:&       LD06& 0\\   
           
       SMP~18     &    \dots     &    11860$^{+800}_{-870}$  &   3590 &   SEA& 0.5\\ 
    
       SMP~20     &    \dots     &    13820$^{+1380}_{-1010}$  &      3880  &   SEA& 1-2\\
    
       SMP~24     &  \dots &  11620$^{+910}_{-740}$  &         2780  &  SEA& 1-2\\
    
       SMP~25     &    \dots & 21100  &  9800  &      LD06& 6-7\\
    
       SMP~26     &   \dots  &   18000 &      300 &     LD06& 8\\
    
       SMP~28     &   \dots  &  19200 &    20700 &     LD06& 8\\

\enddata

\tablenotetext{a}{Temperature for low excitation ions}
\tablenotetext{b}{Adopted by Authors in reference}

\end{deluxetable}

\begin{deluxetable}{lrrrrr}
\tablecaption{Carbon Abundances} 
\tablehead{
\colhead{Name} & \colhead{C$^{+}$/H$^+$} & \colhead{C$^{2+}$/H$^+$} &
\colhead{C$^{3+}$/H$^+$} &  \colhead{ICF(C)}& \colhead{A(C)} \\
}
\startdata

SMP~6& 	1.016$\times10^{-5}$& 1.917$\times10^{-4}$& 2.184$\times10^{-5}$&  1.0& 8.35$^{+0.03}_{-0.03}$ \\ 

SMP~8& 	\dots&                9.596$\times10^{-5}$& 3.631$\times10^{-5}$& 1.0& 8.12$^{+0.17}_{-0.13}$ \\

SMP~13&	2.877$\times10^{-5}$& 4.753$\times10^{-4}$& 3.445$\times10^{-5}$&  1.0& 8.73$^{+0.05}_{-0.15}$ \\

SMP~15& $\ge 4.889\times10^{-6}$& 1.621$\times10^{-4}$& 1.379$\times10^{-5}$& 1.0& 8.26$^{+0.06}_{-0.07}$\\

SMP~16& 7.357$\times10^{-5}$&  7.761$\times10^{-5}$& 4.403$\times10^{-6}$& 1.0& 8.19$^{+0.08}_{-0.1}$\\

SMP~18& $\ge 6.422\times10^{-5}$& 1.377$\times10^{-4}$&  7.047$\times10^{-7}$& 1.0& 8.31$^{+0.05}_{-0.20}$\\

SMP~20& 2.694$\times10^{-5}$& 1.492$\times10^{-4}$&  1.283$\times10^{-6}$& 1.0& 8.25$^{+0.2}_{-0.22}$\\

SMP~24& \dots& 1.485$\times10^{-4}$&  \dots& 1.02& 8.18$^{+0.05}_{-0.25}$\\

SMP~25& & 1.626$\times10^{-6}$&  1.602$\times10^{-6}$&  1.38& 6.64$^{+0.09}_{-0.13}$ \\

SMP~26& \dots& 1.349$\times10^{-4}$&  6.497$\times10^{-5}$&  1.45& 8.46$^{+0.04}_{-0.05}$ \\

SMP~28& \dots& 4.255$\times10^{-6}$&  2.525$\times10^{-6}$&  1.34& 6.96$^{+0.02}_{-0.03}$\\

\enddata

\end{deluxetable}

\end{document}